\documentclass[%
 reprint,
 amsmath,amssymb,
 aps,
]{revtex4-2}

\usepackage{graphicx}
\usepackage{dcolumn}
\usepackage{bm}
\usepackage{float} 
\usepackage{amsmath}
\usepackage{hyperref}
\usepackage{xcolor}
\usepackage[normalem]{ulem}
\definecolor{DarkGreen}{RGB}{50, 175, 50} 

\usepackage{amsmath, amssymb, relsize}
\usepackage{setspace} 
\usepackage{cancel}
\begin{document}

\title{A hybrid method for reconstruction of the equation of state of dark energy and its application to Pantheon+SH0ES data.}

\author{Gawain Simpson} \email{gawain.simpson@utas.edu.au}
\author{Krzysztof Bolejko}  \email{krzysztof.bolejko@utas.edu.au}
\author{Stephen Walters}  \email{stephen.walters@utas.edu.au}
\affiliation{School of Natural Sciences, Sciences and Engineering, University of Tasmania, PO Box 807, TAS 7006, Australia}

\date{\today}

\begin{abstract}
Cosmology aims to understand the physical properties of our Universe on its largest scales. One such feature is the expansion of the Universe, which currently seems to be dominated by a phenomenon referred to as dark energy. The physical nature and properties of dark energy are one of the main topics of investigation of modern cosmology. Observational cosmology aims to reconstruct the evolution and the equation of state of dark energy, while theoretical cosmology aims to provide methods for such reconstructions and models explaining the nature of dark energy. If the equation of state, defined as the ratio of pressure to density $w = p/\rho$, deviates from $-1$, i.e. $w\ne-1$, then this would imply the existence of some sort of dynamical process behind dark energy.
Most investigations assume a specific parametric form of $w$, eg. $w(z) = w_{0} + w_a z/(1+z)$, with $z$ being redshift and $w_0$ and $w_a$ being constants. The analysis of the data is then reduced to fitting the model to the data. 
In this work, we take a different approach. Instead of imposing a predefined parametric form for $w(z)$, we reconstruct the equation of state indirectly from the dimensionless comoving distance $D(z)$ and its derivatives. This avoids assuming a specific physical parametrisation of dark energy, such as the CPL form, but still requires adopting functional representations for the distance--redshift relation itself. The method should therefore be regarded as a hybrid or semi-parametric reconstruction approach: the parametrisation is shifted from the equation of state to the observable distance function.
Finally we apply the method to the Pantheon+ SH0ES data. The results are consistent with dark energy being the cosmological constant, i.e. $w = -1$. Future surveys such as LSST will provide more data and narrow down the uncertainty. This in turn will yield tighter constraints on dark energy.
\end{abstract}

\keywords{Astrophysics - Cosmology and Nongalactic Astrophysics; General Relativity and Quantum Cosmology}

\maketitle

\section{Introduction}

The Universe is expanding --- a discovery first made through the observation that distant galaxies recede from us at speeds roughly proportional to their distances \citep{Hubble168}. Over cosmic time, that expansion has not been steady: it has slowed and accelerated in different epochs as the balance between matter, radiation and dark energy has shifted. Observations of the cosmic microwave background (CMB) show that the Universe expanded much faster at early times, when radiation and matter dominated its energy content \citep{2020AA...641A...6P}. Today, the expansion appears to be accelerating. Type~Ia supernovae at redshifts $z \sim 0.5$--$1$ are dimmer than expected in a decelerating Universe, implying that the cosmic scale factor $a(t)$ has entered an accelerating phase \citep{1998AJ....116.1009R, 1999ApJ...517..565P}.  
\newline
This acceleration requires a component with strongly negative pressure - \textit{dark energy} - which dominates the energy density of the Universe today. In the standard cosmological model ($\Lambda$CDM), dark energy is described by a cosmological constant $\Lambda$ with an equation-of-state parameter $w = p/\rho = -1$. This model has been remarkably successful, but growing observational tensions suggest it may be incomplete. In particular, the \textit{Hubble tension} - a persistent discrepancy between the locally measured expansion rate ($H_0 = 73.30 \pm 1.04 ~\mathrm{km~s^{-1}~Mpc^{-1}}$ \cite{2022ApJ...934L...7R}) and the lower value inferred from the CMB under $\Lambda$CDM ($H_0 = 67.4 \pm 0.5 ~\mathrm{km~s^{-1}~Mpc^{-1}}$ \cite{2020AA...641A...6P}) - hints that the late-time expansion history may not exactly follow the simple cosmological constant model \citep{2019NatAs...3..891V}.  
\newline
One of the most direct ways to explore this possibility is through the \textit{equation of state of dark energy}, $w(z)$. Because the cosmic expansion rate $H(z)$ depends on the energy density and pressure of dark energy, measurements of distances and expansion rates at different redshifts encode information about $w(z)$. Many analyses have attempted to capture this dependence by assuming a specific functional form - for example, a constant $w$, or a two-parameter expansion such as the Chevallier--Polarski--Linder (CPL) form \citep{2001IJMPD..10..213C, 2003PhRvL..90i1301L}.
Such parameterisations are convenient but they inevitably bias the reconstruction toward their chosen form, potentially hiding more complex or time-dependent behaviour, such as might be produced by the evolution of dark energy.
\newline
In this work, we take a different approach. Instead of imposing a predefined model for $w(z)$, we present a \textit{reconstruction} that derives the equation of state directly from observable quantities such as the dimensionless comoving distance $D(z)$ and its derivatives. 
Since the expansion rate $H(z)$ can be expressed in terms of $D'(z)$ and $D''(z)$, the equation of state $w(z)$ can be inferred directly from these quantities without assuming any specific functional form.
However, purely derivative-based reconstructions are known to be sensitive to noise in the data, while fully parametric approaches can bias the result through the assumed form of $w(z)$. To address this, we introduce a hybrid method that combines these approaches: an initial MCMC analysis is used to constrain key cosmological parameters such as $\Omega_M$ and $H_0$, thereby reducing degeneracies, after which $w(z)$ is reconstructed from $D(z)$ and its derivatives. This stabilises the reconstruction while retaining sensitivity to deviations from a cosmological constant. 
While this approach reduces direct model dependence in $w(z)$, the inferred behaviour of the equation of state remains sensitive to the choice of fitting function used to reconstruct $D(z)$, since $w(z)$ depends explicitly on the derivatives $D'(z)$ and $D''(z)$.
This approach therefore provides a controlled balance between parametric bias and non-parametric instability.

The structure of this paper is as follows: Sec.\ref{Sec:Methods} presents the hybrid-method to retrieve the dark energy equation of state; Sec. \ref{Sec:Tests} tests our model against simulated Supernovae data; Sec.~IV provides the application of the methods to the Pantheon+SH0ES data; finally Sec.~V concludes the paper.

\section{Methods}\label{Sec:Methods}

\subsection{Distance and its relation to dark energy}

Within a class of FLRW models, the comoving distance $d(z)$ is defined as 
\begin{equation} \label{eq:Comoving_in_methods}
    d(z) = \frac{c}{H_0}\int_{0}^{z} \frac{ {\rm d} z'}{E(z')},
\end{equation}
where
\begin{equation} \label{eq:Esquared}
    E^2 = \Omega_M(1+z)^3 + \Omega_R(1+z)^4 + \Omega_K(1+z)^2 + \Omega_\Lambda e^{ \left( 3\int {\rm d} z' \frac{1+w}{1+z'}
    \right)},
\end{equation}

and $\Omega_M, \Omega_R, \Omega_K$ and $\Omega_\Lambda$ are the Matter, Radiation, Curvature and Dark Energy Density parameters respectively, $c$ is the speed of light, $H_0$ is the Hubble constant, $z$ is the redshift, and $w$ is the equation of state of dark energy. 

Often the equation of state is parameterized with one or two parameters. For example, the CPL parametrisation is one of most popular parametrisations used, and  has the following form \citep{2001IJMPD..10..213C, 2003PhRvL..90i1301L}:
\begin{equation} \label{eq:CPL}
    w(z) = w_0 + w_a\ \frac{z}{1+z}.
\end{equation}
This is a useful parametrisation for two reasons: (i) it is bounded (i.e. does not diverge as for example $w = w_0 + w_a \cdot (1+z)$), (ii) the sign of $w_a$ provides insight into the behavior of dark energy, i.e. whether its pressure is increasing or decreasing (compared to its energy density) with the cosmic evolution. 

Neglecting radiation and spatial curvature, the distance-redshift relation depends on: $H_0$, $\Omega_M$, $w_0$, and $w_a$ ($\Omega_\Lambda$ is fixed via $\Omega_M + \Omega_\Lambda = 1)$. Often the model contains more parameters, however, this is the minimal cosmological set required to infer evolution of dark energy in a meaningful way. Usually, the analysis is done by the means of MCMC methods.

The drawback of this method is that the dependence on $w$ involves double-integration (firstly to estimate the contribution to energy density to the expansion rate, and secondly, the contribution of the expansion rate to the distance itself). This creates low sensitivity to the variability of $w$ as the variability disappears in the integration. This is also seen in the Taylor expansion of the distance with respect to redshift \citep{2007CQGra..24.5985C}, where the linear term depends on $H_0$, the quadratic term on $H_0$ and $\Omega_m$ and the dependence on $w$ only appears in the cubic coefficient. Consequently, the constraints on $w$ are subject to large uncertainties.

The way around this problem would be to re-arrange eq. (\ref{eq:Comoving_in_methods}) for $w$. Introducing the dimensionless comoving distance D as
\begin{equation}
D(z) = \frac{H_0}{c} d(z),
\end{equation}
and assuming spatial flatness, eq. (\ref{eq:Comoving_in_methods}) can be rearranged to (cf. Appendix \ref{derivation_of_w}) \cite{2007JCAP...08..011C}.
\begin{equation} \label{eq:wfinal_methods}
w = \frac{\frac{2}{3}\frac{D''}{D'}(1+z) + 1}{\Omega_M(1+z)^3{D'}^2 - 1}.
\end{equation}

This reformulation has an important implication. In the standard approach, the dependence of observables on $w(z)$ enters through multiple integrals, which smooth out any time variation and reduce sensitivity to the detailed behaviour of the equation of state. In contrast, eq. (\ref{eq:wfinal_methods}) expresses $w(z)$ in terms of derivatives of the distance-redshift relation, making the reconstruction sensitive to local structure in the data. This has the potential to reveal time variation in $w(z)$ that may otherwise be hidden by the integrated nature of distance measurements.

Despite this advantage, the direct application of eq. (\ref{eq:wfinal_methods}) presents several challenges.
The major obstacle is that we need to know the distance and its derivatives, with respect to $z$. While observations of supernova provide $d_L$, and hence $D$, due to the large spread and uncertainties, direct inference from the data of the value of derivatives $D'$ and $D''$ is problematic. Additionally, eq. (\ref{eq:wfinal_methods}) still requires knowledge about $\Omega_M$ and $H_0$.

A further challenge of this approach is the numerical instability inherent in estimating derivatives from noisy data. The reconstruction of $w(z)$ depends explicitly on both $D'(z)$ and $D''(z)$, making it sensitive to small-scale fluctuations in the inferred distance--redshift relation. In the presence of noise, direct numerical differentiation is ill-conditioned, with small perturbations in $D(z)$ leading to amplified variations in the reconstructed equation of state. To mitigate this, we approximate the data using smooth analytic functions or Gaussian Process reconstructions, which act as implicit regularisation schemes. This stabilises the derivative estimation, but introduces a trade-off between noise suppression and potential bias from the assumed functional form.

In this paper we derive a hybrid method for the inference of $w(z)$ that combines these approaches in a complementary way. Parametric inference is first used to constrain $\Omega_M$ and $H_0$, reducing degeneracies, after which the equation of state is reconstructed from the distance--redshift relation and its derivatives. In this way, the method stabilises the derivative-based reconstruction while avoiding the need to assume a specific functional form for $w(z)$.

\subsection{Simulated Supernovae data}

\begin{figure}[h]
\includegraphics[width=9cm]{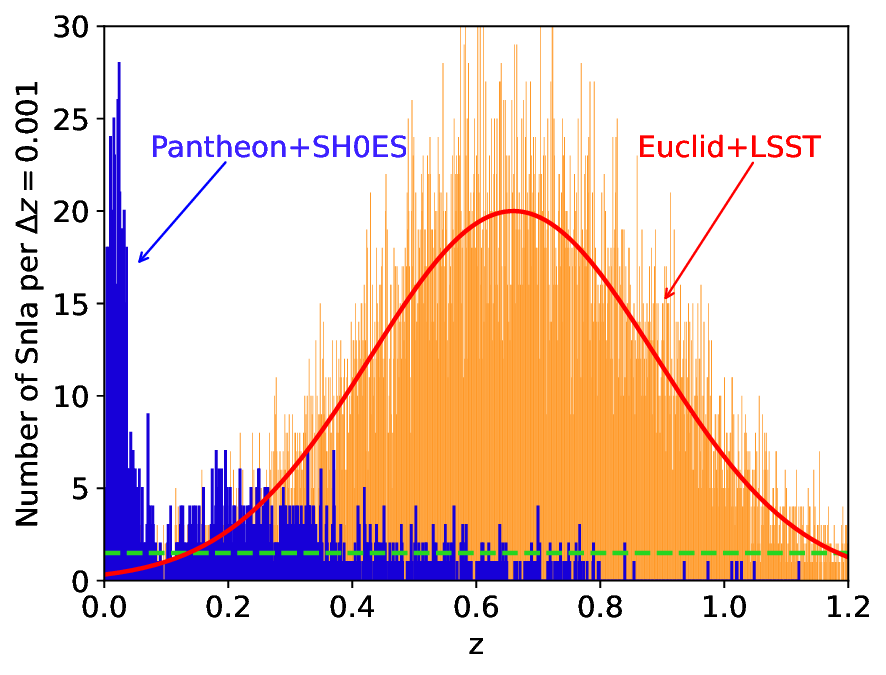}
\caption{Redshift distribution of supernova in the Pantheon+ Dataset \cite{2022ApJ...938..113S} and the expected distribution of Euclid+LSST supernovae \cite{10.1093/mnras/stad2179}. The dashed green curve is a constant curve $n = 1.5 / 0.001 $ (i.e. $1.5$ supernovae per $\Delta z = 0.001$). The solid red curve is the normal distribution with a mean of $\mu = 0.66$ and standard deviation $\sigma = 0.23$. The simulated supernovae data generated based on these distributions is then used to test the methods for $w(z)$ reconstruction.}\label{figure1}
\end{figure}

\begin{figure}[h]
\includegraphics[width=9cm]{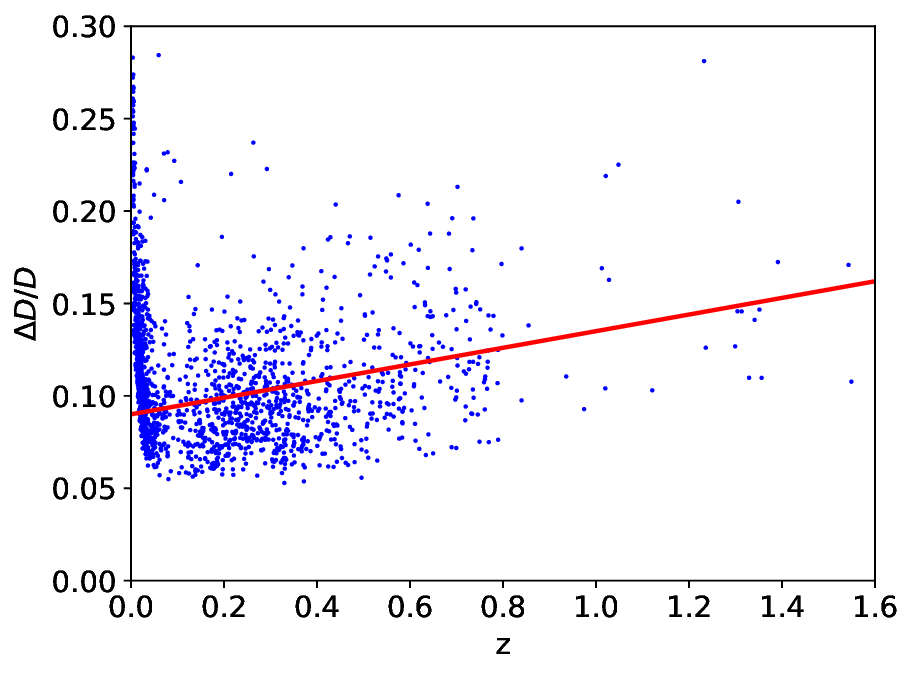}
\caption{Relative uncertainty $\Delta D/D$ of Pantheon+ dataset, where $D$ is defined by eq. (\ref{dimensionless_D}). The solid red line shows the trend of data for $z> 0.01$, given by $(\Delta D/D)_{mean} = 0.09+0.045\,z$.
This data is used to assign uncertainties to the simulated supernovae data. }\label{figure2}
\end{figure}

\begin{figure}[h]
\includegraphics[width=9cm]{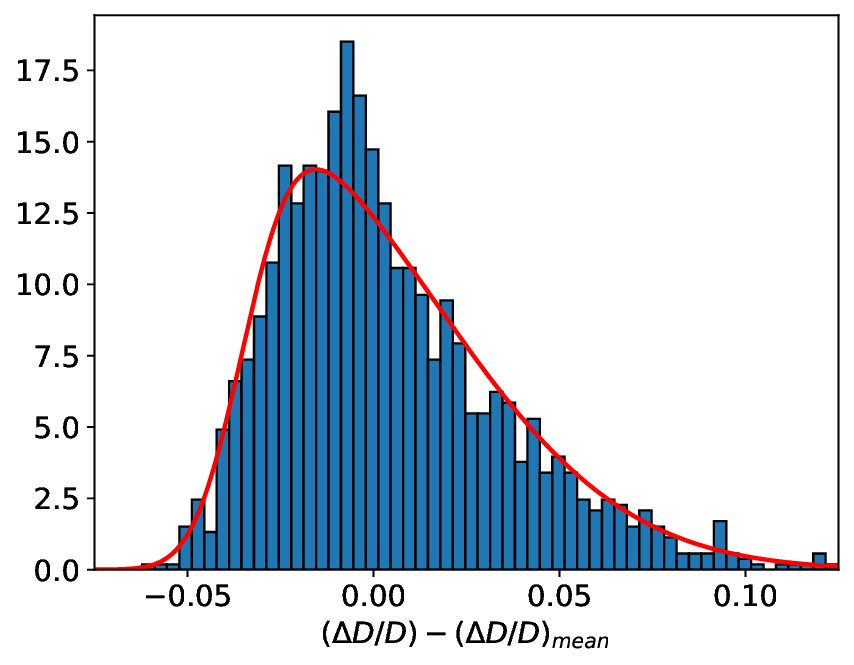}
\caption{The distribution of residuals of relative uncertainties: $\Delta D/D - (\Delta D/D)_{mean}$, cf. Fig. \ref{figure1}.  The solid red curve is the skew normal distribution with the following parameters: location $\xi =-0.035$, scale $\omega = 0.051$ and shape $\alpha = 4.73$. This distribution is used to generate the  data that we use to test our methods for $w(z)$ reconstruction.}\label{figure3}
\end{figure}

In order to demonstrate and test the method we created a set of simulated supernovae data. This simulated data is based on the distance-redshift relation that mimics (to a certain extent) the Pantheon+ Dataset \cite{2022ApJ...938..113S}. The Pantheon+ Dataset contains the light curves of 1701 supernovae. The distribution of the data is presented in Fig. \ref{figure1}. In addition to the Pantheon+ Dataset \cite{2022ApJ...938..113S}, Fig. \ref{figure1} also shows the expected distribution of Euclid+LSST supernovae \cite{10.1093/mnras/stad2179}. 
Thus we will assume that the density of simulated supernovae data is 1.5 per redshift bin of $\Delta z =0.001$. In addition, instead of light curves we will use the dimensionless comoving distance $D(z)$. Firstly, the distance modulus $\mu$ is converted to the luminosity distance $d_L$
\begin{equation} \label{eq:distance_modulus}
\mu = 5 \log_{10} {\left(\frac{d_L}{10\,{\rm pc}}\right)}.
\end{equation} 
Secondly, we assume spatial flatness, meaning the relation between the dimensionless comoving distance and luminosity distance is
\begin{equation}\label{dimensionless_D}
D  = \frac{H_0}{c}\frac{d_L}{1+z}.
\end{equation} 
Each distance is then accompanied with an uncertainty. The uncertainties that mimic those of the Pantheon+ dataset are presented in Figs. \ref{figure2} and \ref{figure3}.

\begin{figure}[h]
\includegraphics[width=9cm]{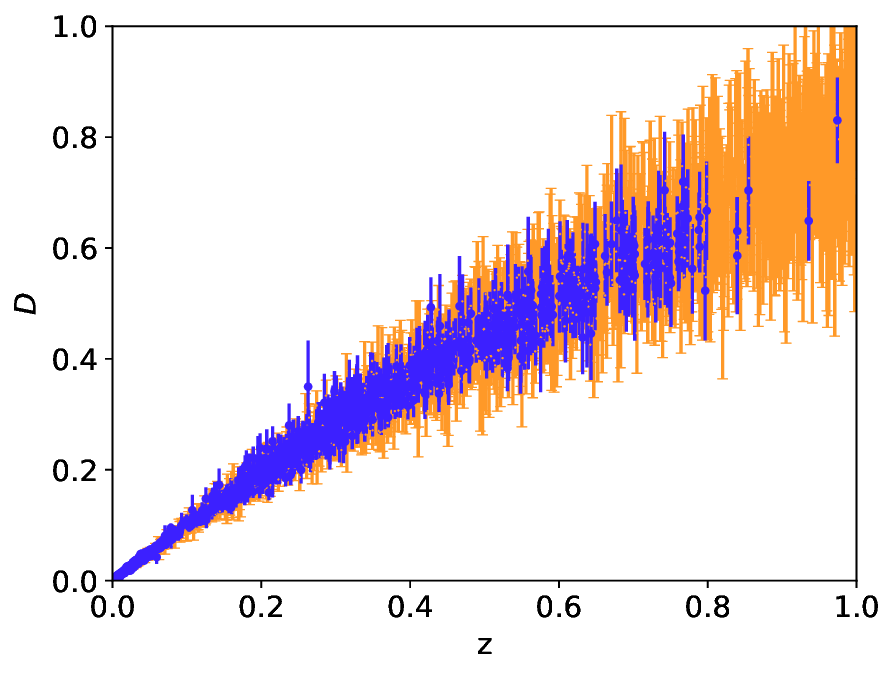}
\caption{The simulated supernovae data of dimensionless comoving distance (light orange) and dimensionless comoving distance of of Pantheon+ supernovae (dark blue). The simulated supernovae data has a constant number density of $n = 1.5$ per $\Delta z = 0.001$, whereas the Pantheon+ supernova follow the distribution presented in Fig. \ref{figure1}. The cosmological parameters used to generate this particular simulated supernovae data are: $H_0 = 74 \,{\rm km} \,{\rm s}^{-1} \,{\rm Mpc}^{-1}$, $\Omega_M = 0.3$, $\Omega_\Lambda = 0.7$, $w_0 = -1.1$, and $w_a = 0.2$.}\label{figure4}
\end{figure}

Thus, the simulated supernovae catalog is created as follows: (i) a redshift array with number density of $n = 1.5$ supernova per redshift bin of $z =0.001$ is generated; (ii) for each of these redshifts the dimensionless distance $D_0$ is evaluated; (iii) the mean relative uncertainty is then calculated $(\Delta D/D)_{mean} = 0.09+0.045\,z$; \newline
(iv) The adopted uncertainty model is constructed empirically
to reproduce the redshift-dependent scatter observed in the
Pantheon+ sample. The functional form $(\Delta D/D)_{mean} = 0.09+0.045\,z$
is obtained from a fit to the trend of relative uncertainties in the Pantheon+ data (cf. Fig.~2), while the residual scatter about this trend is modelled using a skew-normal distribution chosen to reproduce the observed asymmetry in the error distribution (cf. Fig.~3). This approach is not intended to provide a full statistical description of the Pantheon+ covariance, but rather to generate mock datasets with realistic noise amplitude and redshift dependence for the purpose of testing the reconstruction method.
(v) We note that this simplified noise model does not incorporate the full covariance structure of the Pantheon+ dataset, which includes correlated systematic uncertainties between supernovae. Incorporating the full covariance matrix would provide a more complete statistical description, but is not required for the present tests, which are designed to assess the stability and bias of the reconstruction under realistic noise levels rather than to reproduce the exact statistical properties of the Pantheon+ sample.
(vi) the relative uncertainty is then adjusted by a random number drawn from a skew normal distribution with location $\xi =-0.035$, scale $\omega = 0.051$ and shape $\alpha = 4.73$; (vii) finally, the 
dimensionless distance $D$ is adjusted by taking a random number drawn from a normal distribution with the mean $D_0$ and the standard deviation $\sigma = (\Delta D/D)$ as adjusted according to the previous point (vi).  An example of simulated supernovae data generated in this way is presented in Fig. \ref{figure4}.

\subsection{Reconstruction of the equation of state of dark energy }

\begin{figure}[h]
\includegraphics[width=9cm]{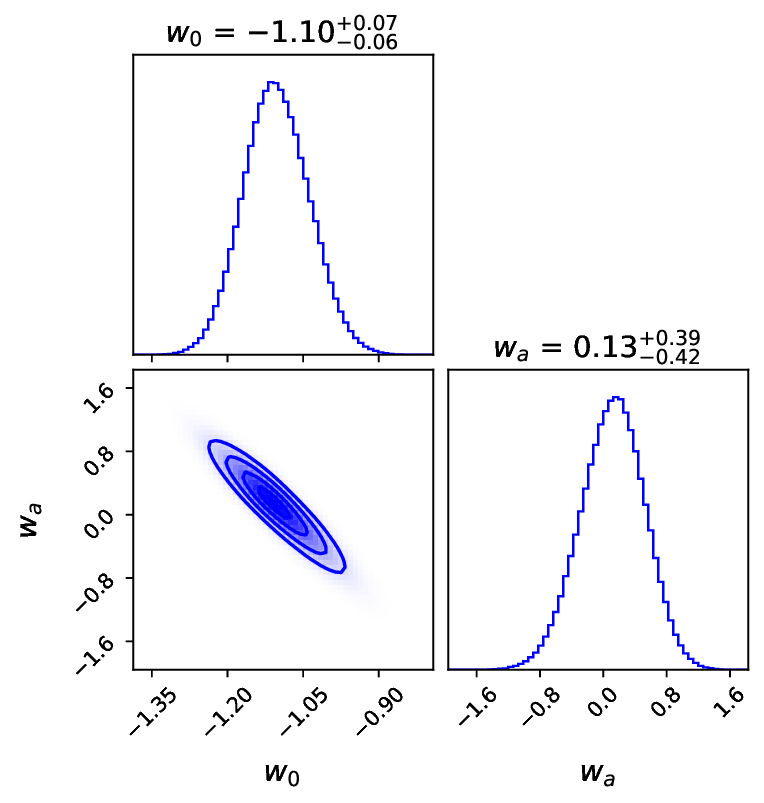}
\caption{Constraints on the CPL parameters $w_0$ and $w_a$ using the simulated supernovae with redshift number density of $n = 1.5$ per $\Delta z = 0.001$, and keeping other cosmological parameters fixed. The cosmological parameters used to generate the simulated supernovae data are: $H_0 = 74 \,{\rm km} \,{\rm s}^{-1} \,{\rm Mpc}^{-1}$, $\Omega_M = 0.3$, $\Omega_\Lambda = 0.7$, $w_0 = -1.1$, and $w_a = 0.2$.}\label{figure5}
\end{figure}

\begin{figure}[h]
\includegraphics[width=9cm]{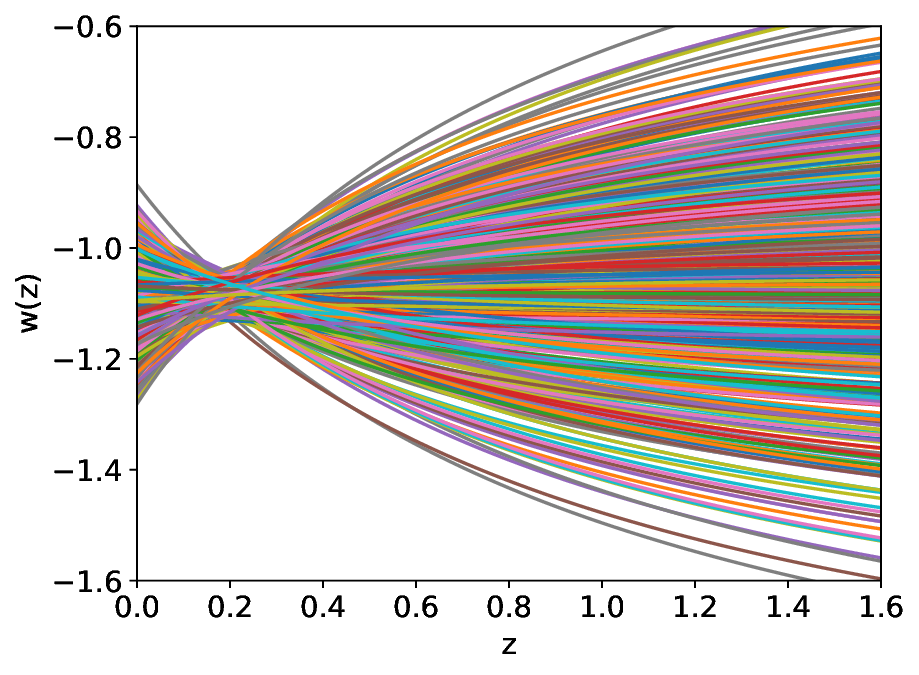}
\caption{Reconstruction of the equation of state of dark energy $w(z)$. When using the MCMC analysis, the parameters $w_0$ and $w_a$ are probed. These are then used with eq. (\ref{eq:CPL}) to reconstruct $w(z)$. For clarity of the figure only 640 lines (out of 800,000) are plotted.}\label{figure6}
\end{figure}

\begin{figure}[h]
\includegraphics[width=9cm]{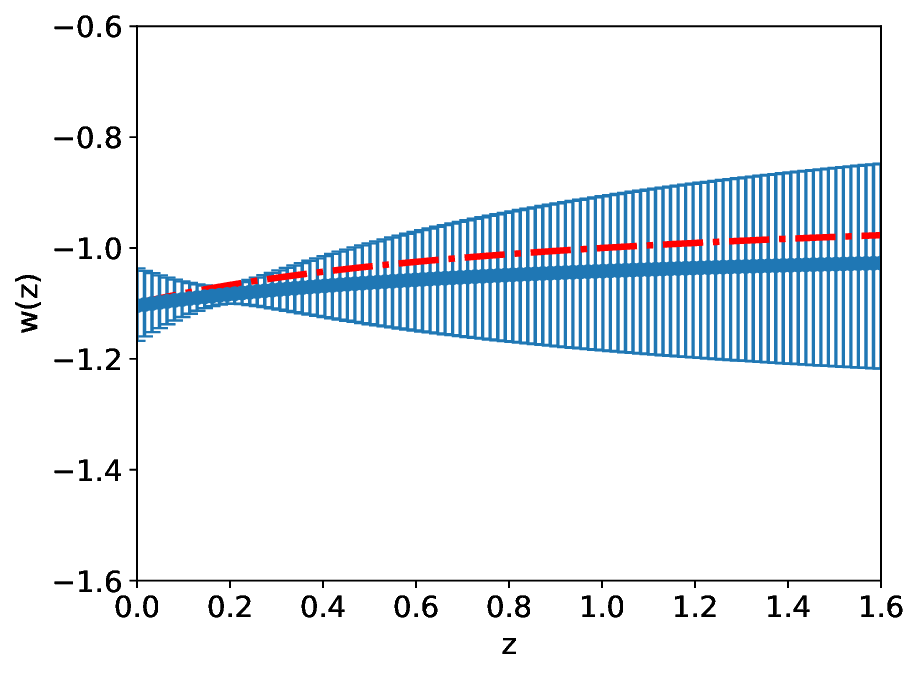}
\caption{Reconstruction of the equation of state of dark energy $w(z)$. Similar to Fig. \ref{figure6} but only the interval containing 68\% of all 800,000 curves is plotted. The dash-dot red line shows the actual $w(z)$ based on the $w_0 = -1.1$ and $w_a = 0.2$, which was used to generate the simulated supernovae data.}\label{figure7}
\end{figure}

\begin{figure}[h]
\includegraphics[width=9cm]{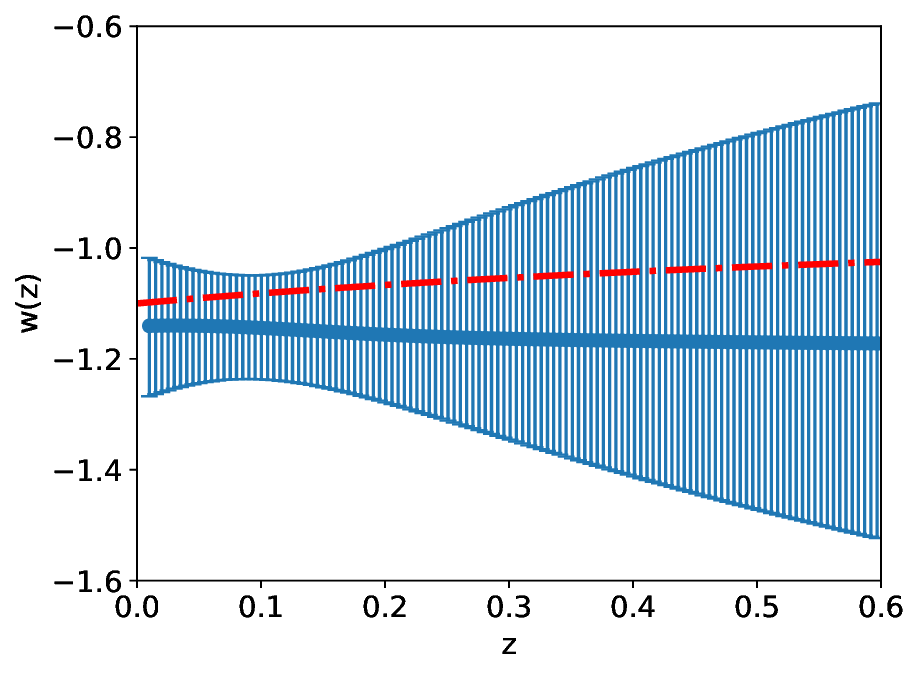}
\caption{Reconstruction of the equation of state of dark energy $w(z)$. Similar to Fig. \ref{figure7} but this time $H_0$ and $\Omega_M$ are also explored by the MCMC analysis. Also, the horizontal axis (redshift range) has been reduced to [0:0.6]. The dash-dot red line shows the actual $w(z)$ based on the $w_0 = -1.1$ and $w_a = 0.2$, which was used to generate the simulated supernovae data.}\label{figure8}
\end{figure}

The purpose of this section is to test and discuss the method of reconstruction of the equation of state of dark energy $w(z)$ before applying it to real data. 
We use a model with dark energy parameterized with the CPL model, eq. (\ref{eq:CPL}), and the following parameters: $H_0 = 74 \,{\rm km} \,{\rm s}^{-1} \,{\rm Mpc}^{-1}$, $\Omega_M = 0.3$, $\Omega_\Lambda = 0.7$, $w_0 = -1.1$, and $w_a = 0.2$. This background model is then used to generate the mock data. 
This allows a direct comparison between the reconstructed equation of state and the underlying model used to generate the data, providing a controlled test of the method's accuracy.
The simulated supernovae data is generated as described in the previous section with the supernovae redshift number density of $n = 1.5$ per $\Delta z = 0.001$, and $z<1.6$. We then use the code \texttt{emcee} \footnote{\url{https://emcee.readthedocs.io/en/stable/}} \cite{2013PASP..125..306F} to infer the parameters $w_0$ and $w_a$. The results are presented in Figs. \ref{figure5}--\ref{figure7}. Due to the nature of the MCMC analysis that probes $w_0$ and $w_a$ it appears that the greatest sensitivity to detect any deviation of dark energy from the cosmological constant scenario (i.e. $w(z) \ne -1$) is to probe the low-redshift Universe about  $z \approx 0.2$.  Figure \ref{figure7} seems to suggest that the uncertainty in this region could be very small. However, if one allows for other cosmological parameters to vary, instead of keeping them fixed then the uncertainty becomes larger. This is presented in Fig. \ref{figure8}, which is similar to Fig. \ref{figure7}, but this time $H_0$ and $\Omega_M$ are also explored by the MCMC analysis.
This time, unsurprisingly, the equation of state of dark energy is subject to larger uncertainties, as the number of free-parameters has increased.

\begin{figure}[h]
\includegraphics[width=9cm]{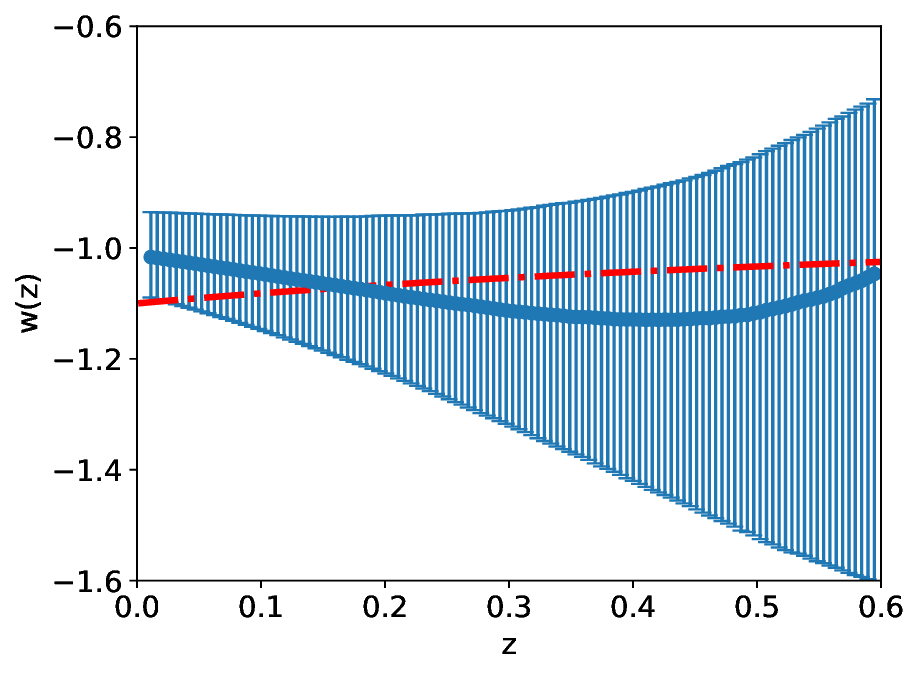}
\caption{Reconstruction of the equation of state of dark energy $w(z)$. Similar to Fig. \ref{figure8} but the equation of state was inferred using eq. (\ref{eq:wfinal_methods}). The dash-dot red line shows the actual $w(z)$ based on the $w_0 = -1.1$ and $w_a = 0.2$, which was used to generate the simulated supernovae data.}\label{figure9}
\end{figure}

If instead of fitting a parametric form of $w$, we use eq. (\ref{eq:wfinal_methods}) then we could in principle reduce the number of free-parameters and consequently keep the uncertainties under control. This however is more difficult to implement than it initially sounds. First we do require the derivatives of $D$. Given large uncertainties in the data we try to overcome this by fitting a polynomial to the distance data instead. As a first choice we use a constrained quadratic polynomial, i.e. $D_0 =0$, and $D_1 = 1$ (cf. \cite{2007CQGra..24.5985C}). Secondly, eq.  (\ref{eq:wfinal_methods}) requires knowledge of $\Omega_M$ and $H_0$. To overcome this hurdle we use the MCMC chains for $\Omega_M$ and $H_0$ from the previous method. The outcome is presented in Fig. \ref{figure9}.

While the uncertainties and sensitivity to reconstruct $w(z)$ is not diametrically different than the standard MCMC methods, it provides a different way of inferring the equation of state, and thus prompts further investigation, which we provide in the next section.

\section{Tests}\label{Sec:Tests}
The use of analytic functions to reconstruct $D(z)$ introduces an additional source of systematic uncertainty, as the choice of fitting function can bias the inferred behaviour of $w(z)$. Since the reconstruction depends on derivatives of the fitted function, any mismatch between the true distance-redshift relation and the adopted functional form propagates non-linearly into the recovered equation of state. In this section, we therefore interpret the performance of different fitting functions in terms of a bias--variance trade-off: simpler functions may stabilise the reconstruction but introduce bias through insufficient flexibility, while more complex functions reduce bias at the cost of increased variance and sensitivity to noise.

The main hurdle in applying eq. (\ref{eq:wfinal_methods}) are derivatives of $D$.
By fitting a smooth function to the comoving distance data, and taking its first and second derivatives, it is possible to retrieve the equation of state of dark energy $w$ from the data. Uncertainties in $w$ can also be calculated by propagating the covariance matrix of any fitted function through the derivatives of the function, plus the uncertainty in $\Omega_M$, the matter density parameter.

In this section we test the following forms of the distance-redshift relation
\begin{align}
\text{Quadratic:~}& D = a_0 z^2 + a_1 z + a_2, \label{eq:Quadratic}\\
\text{Cubic:~}& D = a_0 z^3 + a_1 z^2 + a_2 z + a_3,\label{eq:Cubic}\\
\text{Quartic:~}& D = a_0 z^4 + a_1 z^3 + a_2 z^2 + a_3 z + a_4, \label{eq:Quartic}\\
\text{Exponential:~}& D = a_0 z e^{bz}, \label{eq:Exponential}\\
\text{Logarithmic:~}& D = a_0\, log(1+ a_1 z) + a_2, \label{eq:Logarithmic}\\
\text{Rational fraction:~}& D = \frac{a_0 z + a_1 z^2}{1 + a_2 z + a_3 z^2}. \label{eq:Rational_fraction}
\end{align}

\begin{figure}[h]
\includegraphics[width=9cm]{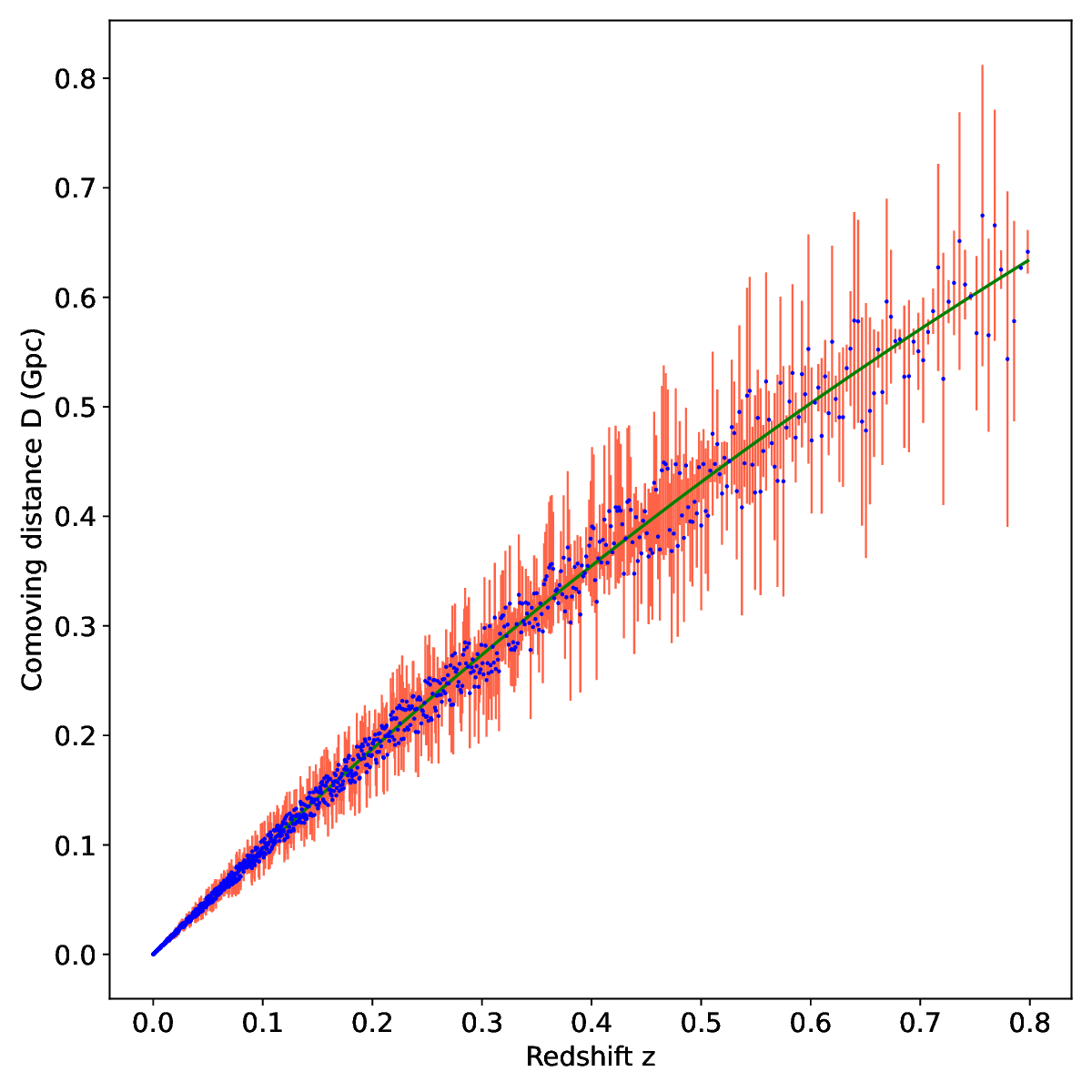}
\caption{\label{fig:Generated_SN} Comoving distance as a function of redshift for 1,000 simulated supernovae, with uncertainties derived from a fit to the Pantheon+ error distribution. Distances are computed within the FLRW framework, and the green curve indicates an exponential model fitted to the data points.}
\end{figure}

In order to test the utility of  eq. (\ref{eq:wfinal_methods}), we generated a number of supernovae. The generated supernovae were distributed with a redshift weighting that mimics the low-z dominance of the Pantheon+ sample.
Figure \ref{fig:Generated_SN} shows a plot of the comoving distance for 1,000 generated SN with uncertainties, using the FLRW metric to calculate the comoving distance. The uncertainties are generated from a curve fitted to the uncertainties in the Pantheon+ data and the green curve is an exponential function fitted to the points. The plot only shows 1,000 SN for clarity, rather than the 10,000 used for the majority of the fitting of the functions to the data.

Using the CPL form of $w$  (eq. \ref{eq:CPL}) to vary the equation of state with $w_0$ from -1.5 to -0.5 and $w_a$ from -1 to 1, both with steps of 0.25, we generated between $10^3$ and $10^6$ SN. We then fitted quadratic, cubic, quartic, exponential, logarithmic and rational fraction curves to each set of SN and used the derivatives of the equations of these fitted curves to calculate the equation of state $w$ using equation (\ref{eq:wfinal_methods}) to attempt to recover $w$. The different quantities of SN generated were to determine how many were necessary to recover $w$ effectively. The parameters used to generate the simulated supernovae data were $H_0 = 73.30 \pm 1.04 ~\mathrm{km~s^{-1}~Mpc^{-1}}$ \cite{2022ApJ...934L...7R}, $\Omega_M = 0.334 \pm 0.018$ \cite{2022ApJ...938..113S} and $\Omega_\Lambda = 1-\Omega_M$.

Figures \ref{fig:w_from_Quadratic_generated_SN} to \ref{fig:w_from_Rational_Fraction_generated_SN} in show the derived equation of state $w$ for curves fitted to 10,000 generated Supernovae for redshift $z$ from 0 to 0.8. The curves fitted to the data are quadratic, cubic, quartic, exponential, logarithmic and rational fraction, respectively. Each shows the equation of state $w$ obtained from fitting the curve and applying equation (\ref{eq:wfinal_methods}), along with their uncertainties. In each case, the uncertainty in $\Omega_M$ was fully propagated through the calculation of $w$ and its uncertainites. The black curve in each figure is $w$ from the CPL equation (eq. \ref{eq:CPL}), which the curve fitting is trying to recover.

The performance of the different fitting functions shows a clear dependence on their flexibility and ability to capture the curvature of the distance--redshift relation.
For low-order polynomials, the quadratic fit (Fig. \ref{fig:w_from_Quadratic_generated_SN}) recovers $w$ accurately only when $w_0 = -1$ and only up to $z = 0.4$; for other values it fails to reproduce the underlying behaviour.
Similarly, the cubic (Fig. \ref{fig:w_from_Cubic_generated_SN}) provides a slightly better reconstruction but only up to $z = 0.4$ in most cases. Only for a very few combinations of $w_0$ and $w_a$ is there a good fit up to $z = 0.8$, such as for $w_0 = -1, w_a = -0.75$ or $w_0 =-0.75, w_a = 0.5$.

For higher-order polynomials, the quartic fit (Fig. \ref{fig:w_from_Quartic_generated_SN}) performs significantly better
for all values of $w_0$ and $w_a$ up to $z=0.4$ and up to $z= 0.8$ in some cases. It is a likely candidate for applying to the Pantheon+ data set.

Among the analytic forms, the exponential (Fig. \ref{fig:w_from_Exponential_generated_SN}) and logarithmic (Fig. \ref{fig:w_from_Logarithmic_generated_SN}) fits generally provide poor reconstructions of $w(z)$. The exponential fit performs well only in limited cases where $w_0 = -1$, while the logarithmic form provides a reasonable fit only for a small subset of models, typically where $w_0 = -0.75$ or $-0.5$ and $w_a < 0$, and therefore neither are robust reconstruction methods.

The rational fraction (Fig. \ref{fig:w_from_Rational_Fraction_generated_SN}) provides the most accurate recovery of $w$ 
across the range of $w_0$ and $w_a$, fitting well up to $z=0.8$ in most cases, but with larger uncertainties than other fits. The larger uncertainties are due to the more complicated rational fraction equation allowing more freedom when fitted to the data. The rational fraction is also a likely candidate for applying to the Pantheon+ data set.

The behaviour of the different fitting functions can be understood in terms of how well they capture the curvature of the distance-redshift relation. Since the reconstruction of $w(z)$ depends explicitly on the second derivative $D''(z)$, any mismatch in the curvature of the fitted function leads directly to errors in the inferred equation of state. Low-order polynomials lack sufficient flexibility to reproduce the underlying curvature over an extended redshift range, resulting in biased reconstructions. Higher-order and more flexible forms, such as quartic polynomials and rational functions, are better able to capture this behaviour, but at the cost of increased variance and larger propagated uncertainties due to the greater number of free parameters.

This trade-off between bias and variance is a key factor in determining the effectiveness of any reconstruction method based on derivatives of observational data.
{In the context of equation (5), this trade-off directly determines the accuracy of the reconstructed equation of state, as biases in the curvature of $D(z)$ translate into systematic errors in $w(z)$, while increased variance leads to amplified fluctuations in the derivative estimates.

An important limitation of the present approach is that, although no explicit parametrisation of $w(z)$ is imposed, the reconstruction is not fully non-parametric. The inferred equation of state depends on the assumed functional form used to fit the distance--redshift relation $D(z)$. Since equation (5) involves the first and second derivatives of $D(z)$, the analytic structure of the chosen fitting function directly constrains the space of allowed behaviour in the reconstructed $w(z)$. In this sense, the parametrisation is shifted from the dark energy equation of state itself to the observable distance ansatz.

This dependence is evident in the differing behaviour obtained from quadratic, quartic, rational and Gaussian Process reconstructions. Simpler fitting functions can bias the reconstruction through insufficient flexibility, while more flexible forms reduce bias at the cost of increased variance and uncertainty propagation. The method should therefore be interpreted as a semi-parametric reconstruction technique, rather than a fully model-independent determination of $w(z)$.

Figures \ref{fig:w_from_Quartic_1000_generated_SN} and \ref{fig:w_from_Quartic_1000000_generated_SN} are quartic fits to 1000 and 1,000,000 SN respectively. Between them they show the effect of varying numbers of SN. For 1000 SN, the uncertainties are large, thus a large range of $w$ is possible. The much larger error bars shows the increased freedom of the fit to the data from a much smaller data set. 
By comparison, the uncertainties for 1,000,000 SN are only a little smaller than for 10,000 SN, indicating little advantage gained by increasing the number of SN above 10,000. Beyond this point, the reconstruction exhibits a clear saturation in performance, indicating that further increases in sample size do not lead to commensurate gains in accuracy.

An important feature of these results is that increasing the number of supernovae beyond $\sim 10^4$ yields only limited improvement in the recovery of $w(z)$. This suggests that, for the class of reconstruction methods considered here, uncertainties are not solely driven by statistical noise, but are increasingly dominated by the choice of reconstruction model and the stability of derivative estimation. In other words, once the data reach a sufficient density, systematic effects associated with the functional form of $D(z)$ and the propagation of noise through derivatives become the limiting factors. This highlights the importance of model selection and regularisation in future high-precision datasets.

Finally, we also tested a second variation of the equation of state $w$, which follows a sinusoidal form 
\begin{equation} \label{eq:sine_variable_w}
w = w_{_0} + w_{_1}sin({w_{_2}}z),
\end{equation}
where $w_0$ and $w_1$ are variable and $w_2 = 6\pi$.
This form is to provide a more difficult test for our approach. The results of this test are shown in figures \ref{fig:sin_w_from_Quartic_generated_SN} and \ref{fig:sin_w_from_Rational_Fraction_generated_SN}.
As seen, tests of the sine form of $w$ for 10,000 SN with a quartic fit (Fig. \ref{fig:sin_w_from_Quartic_generated_SN}) and a rational fraction fit (Fig. \ref{fig:sin_w_from_Rational_Fraction_generated_SN}) show that our model is not flexible enough to recover $w$ for such a widely varying equation of state.
This limitation reflects the smoothing inherent in distance-based observables, which suppresses sensitivity to rapidly varying features in the equation of state. As such, this behaviour is not specific to the method presented here but rather highlights a more general limitation of reconstruction approaches based on integrated quantities such as the luminosity distance.

One concern of using equation (\ref{eq:wfinal_methods}) is whether the denominator remains safely away from zero over the explored redshift range, i.e. $\Omega_M(1+z)^3{D'}^2 - 1$ does not approach zero. Figure \ref{fig:Denominator_of_all_curves} shows the denominator of all curves fitted to the Pantheon+ data. None of the denominators approach zero over the redshift range explored in this paper.

Despite these differences, the main qualitative result, that the reconstructed equation of state is consistent with $w \approx -1$ within uncertainties, remains stable across the range of fitting functions considered. This suggests that, while the detailed behaviour of $w(z)$ is sensitive to the choice of reconstruction method, the overall conclusion is robust to reasonable variations in the functional ansatz used to model $D(z)$.

Because the reconstruction of $w(z)$ depends on derivatives of the distance-redshift relation, it is inherently sensitive to noise in the  data. We therefore emphasise that the purpose of the mock datasets is to probe the stability of the reconstruction under varying noise levels and sampling densities, rather than to reproduce a unique noise realisation. The results presented in this section, including the dependence on the number of supernovae (Figs. 18–19) and the comparison between different fitting functions, demonstrate that the dominant source of variation in the reconstructed $w(z)$ arises from the interplay between noise and model flexibility, rather than the precise form of the adopted uncertainty distribution. This suggests that the qualitative conclusions regarding the recoverability of $w(z)$ are robust to reasonable variations in the noise prescription.

\begin{figure}[h]
\includegraphics[width=9cm]{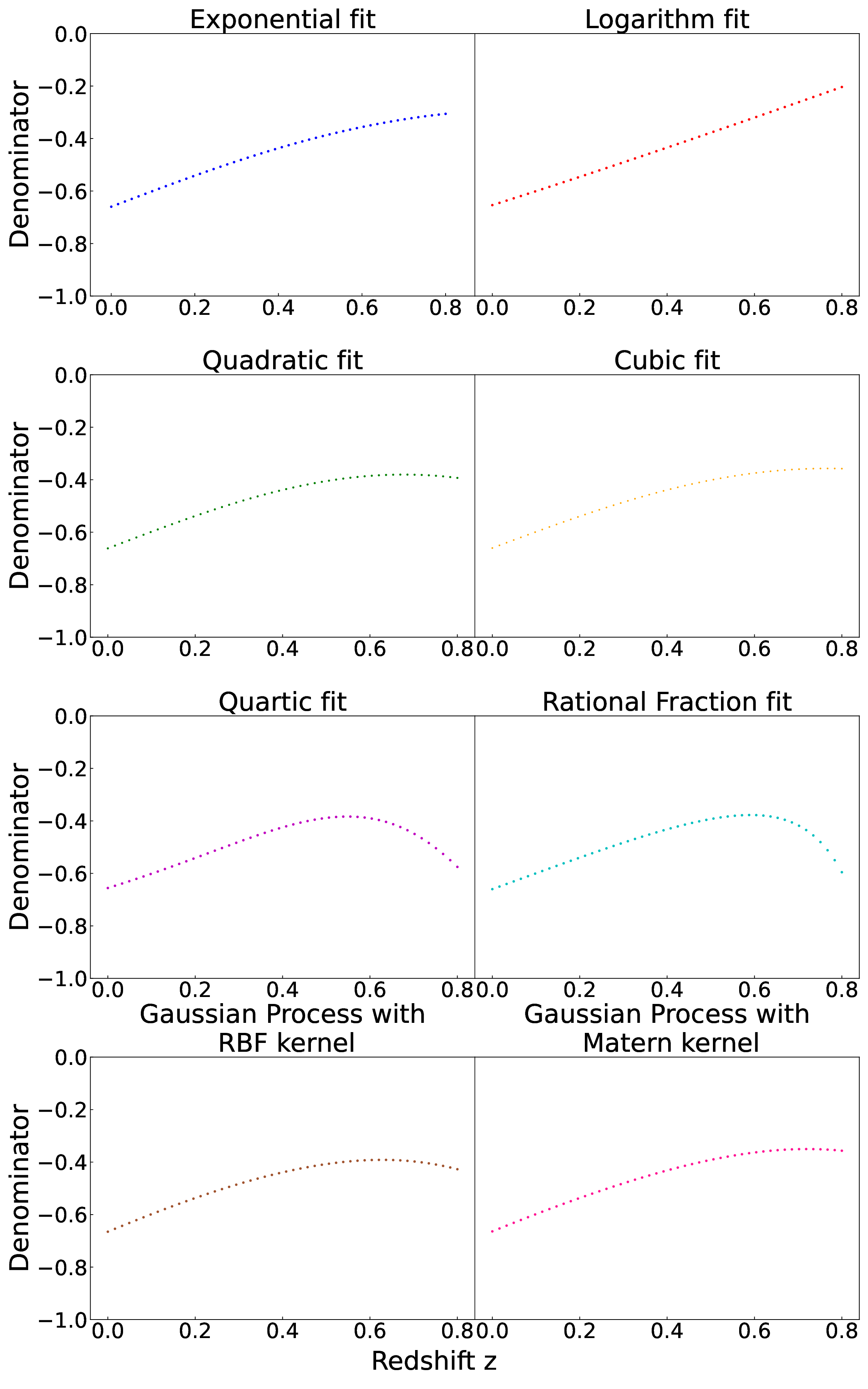}
\caption{\label{fig:Denominator_of_all_curves}Grid of the denominator of all curves fitted to the Pantheon+ data.}
\end{figure}

\begin{figure*}
\includegraphics[height=22.5cm]{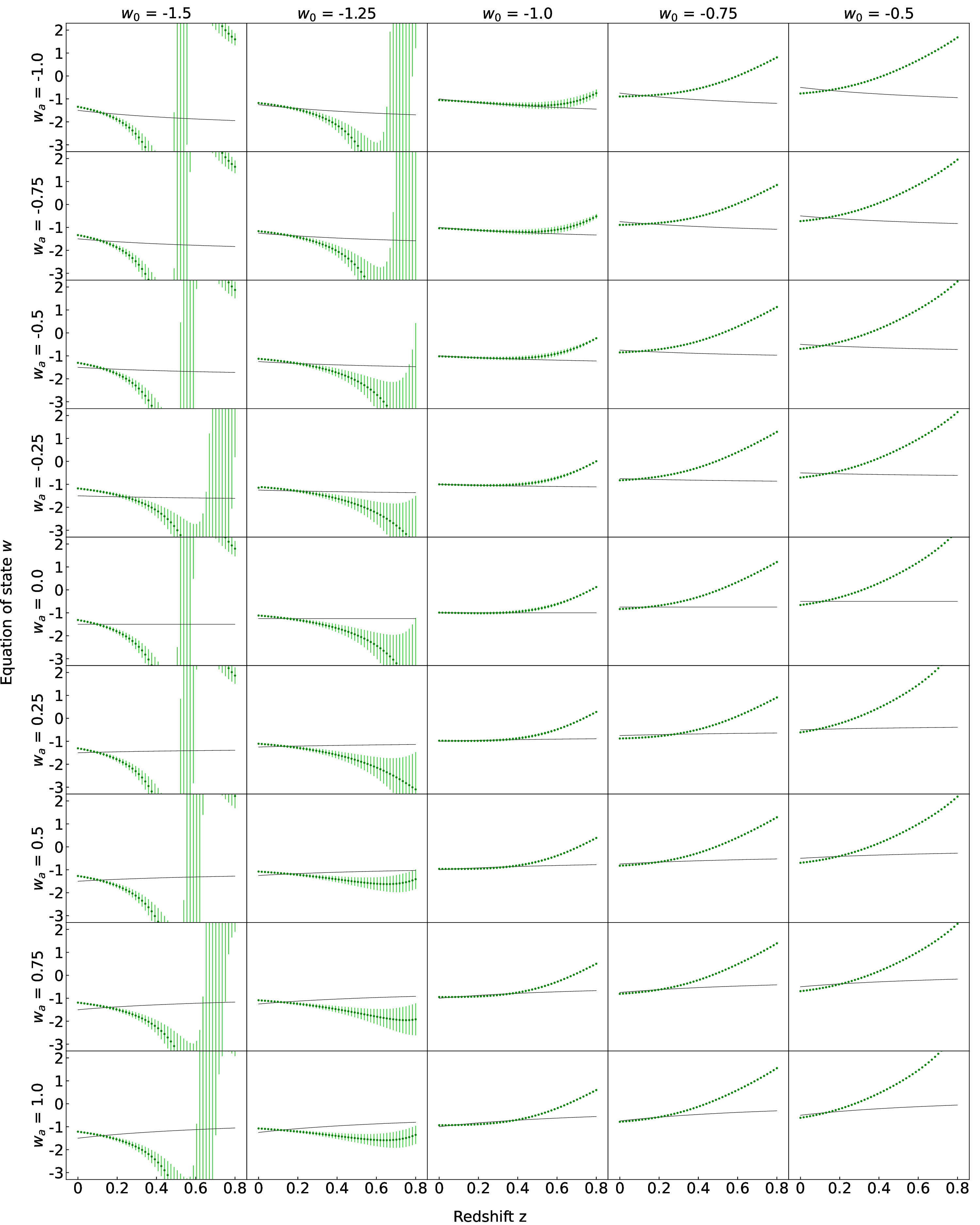}
\caption{\label{fig:w_from_Quadratic_generated_SN}Grid of equation of state values $w(z)$ for 10,000 simulated supernovae, where each $w$ is computed using equation \ref{eq:wfinal_methods} applied to a Quadratic fit (eq. \ref{eq:Quadratic}) to the generated SN data. The plots represent a variety of values of $w_0$ and $w_a$ (CPL equation \ref{eq:CPL}) from -1.5 to -0.5 horizontally, and -1 to 1 vertically, both with a step of 0.25. The black curve represents $w$ from the CPL equation (eq. \ref{eq:CPL}) for the same values. The parameters used to generate the simulated SN data were $H_0 = 73.30 \pm 1.04 ~\mathrm{km~s^{-1}~Mpc^{-1}}$ \cite{2022ApJ...934L...7R}, $\Omega_M = 0.334 \pm 0.018$ \cite{2022ApJ...938..113S} and $\Omega_\Lambda = 1-\Omega_M$.}
\end{figure*}

\begin{figure*}
\includegraphics[height=22.5cm]{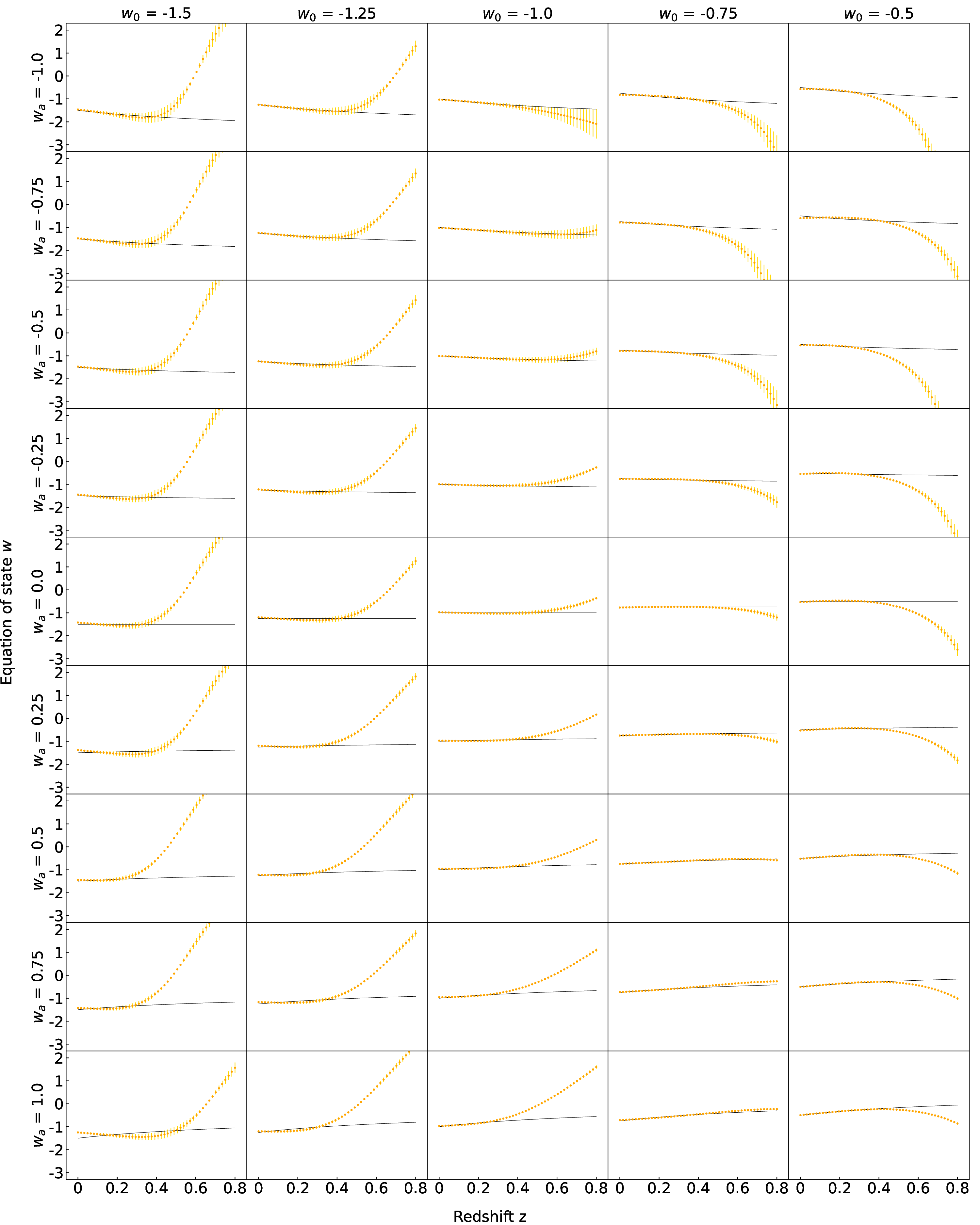}
\caption{\label{fig:w_from_Cubic_generated_SN}Grid of equation of state values $w(z)$ for 10,000 simulated supernovae, where each $w$ is computed using equation \ref{eq:wfinal_methods} applied to a Cubic fit (eq. \ref{eq:Cubic}) to the generated SN data. The plots represent a variety of values of $w_0$ and $w_a$ (CPL equation \ref{eq:CPL}) from -1.5 to -0.5 horizontally, and -1 to 1 vertically, both with a step of 0.25. The black curve represents $w$ from the CPL equation (eq. \ref{eq:CPL}) for the same values. The parameters used to generate the simulated SN were the same as for Fig. \ref{fig:w_from_Quadratic_generated_SN}.}
 
\end{figure*}

\begin{figure*}
\includegraphics[height=22.5cm]{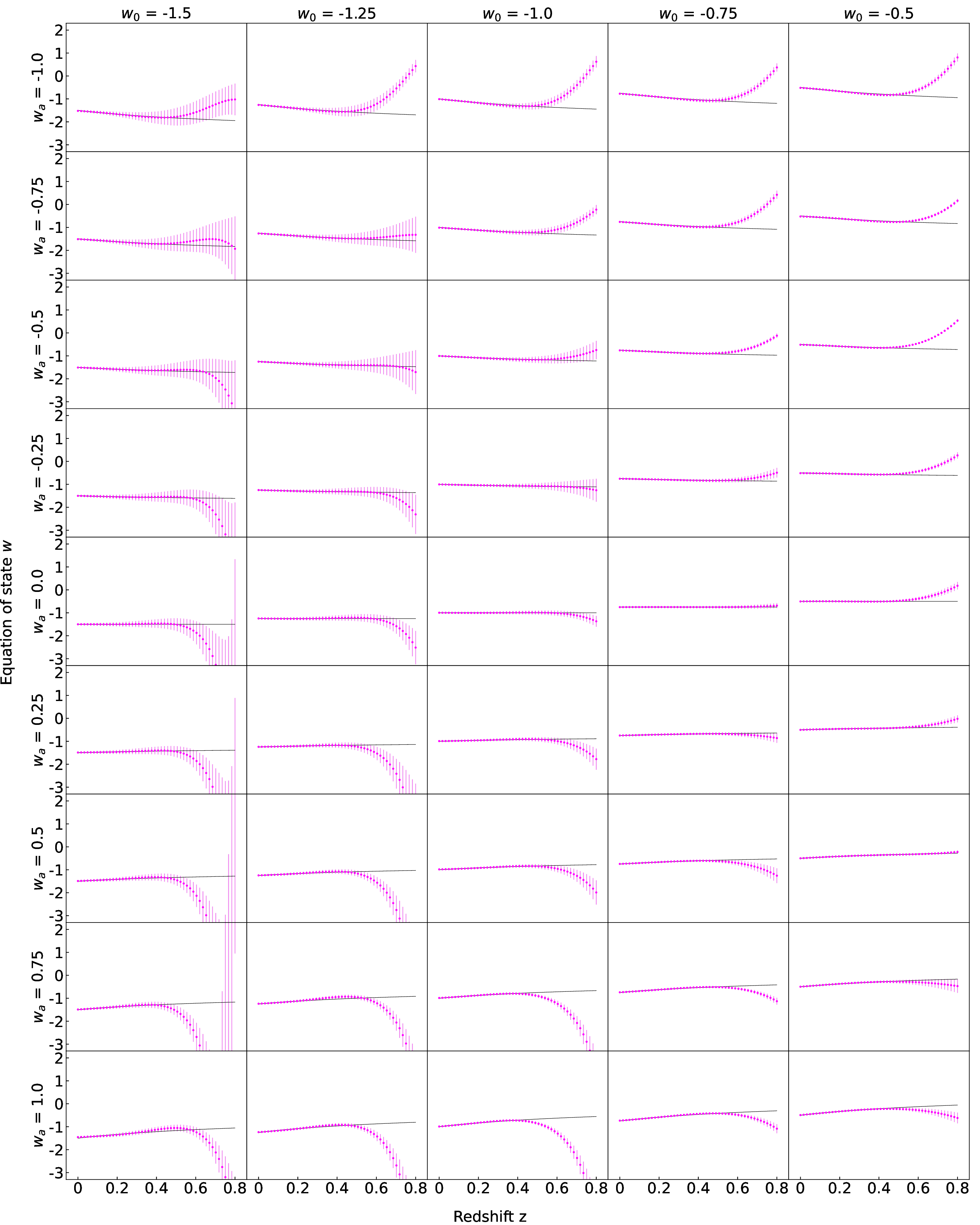}
\caption{\label{fig:w_from_Quartic_generated_SN}Grid of equation of state values $w(z)$ for 10,000 simulated supernovae, where each $w$ is computed using equation \ref{eq:wfinal_methods} applied to a Quartic fit (eq. \ref{eq:Quartic}) to the generated SN data. The plots represent a variety of values of $w_0$ and $w_a$ (CPL equation \ref{eq:CPL}) from -1.5 to -0.5 horizontally, and -1 to 1 vertically, both with a step of 0.25. The black curve represents $w$ from the CPL equation (eq. \ref{eq:CPL}) for the same values. The parameters used to generate the simulated SN were the same as for Fig. \ref{fig:w_from_Quadratic_generated_SN}.}
\end{figure*}

\begin{figure*}
\includegraphics[height=22.5cm]{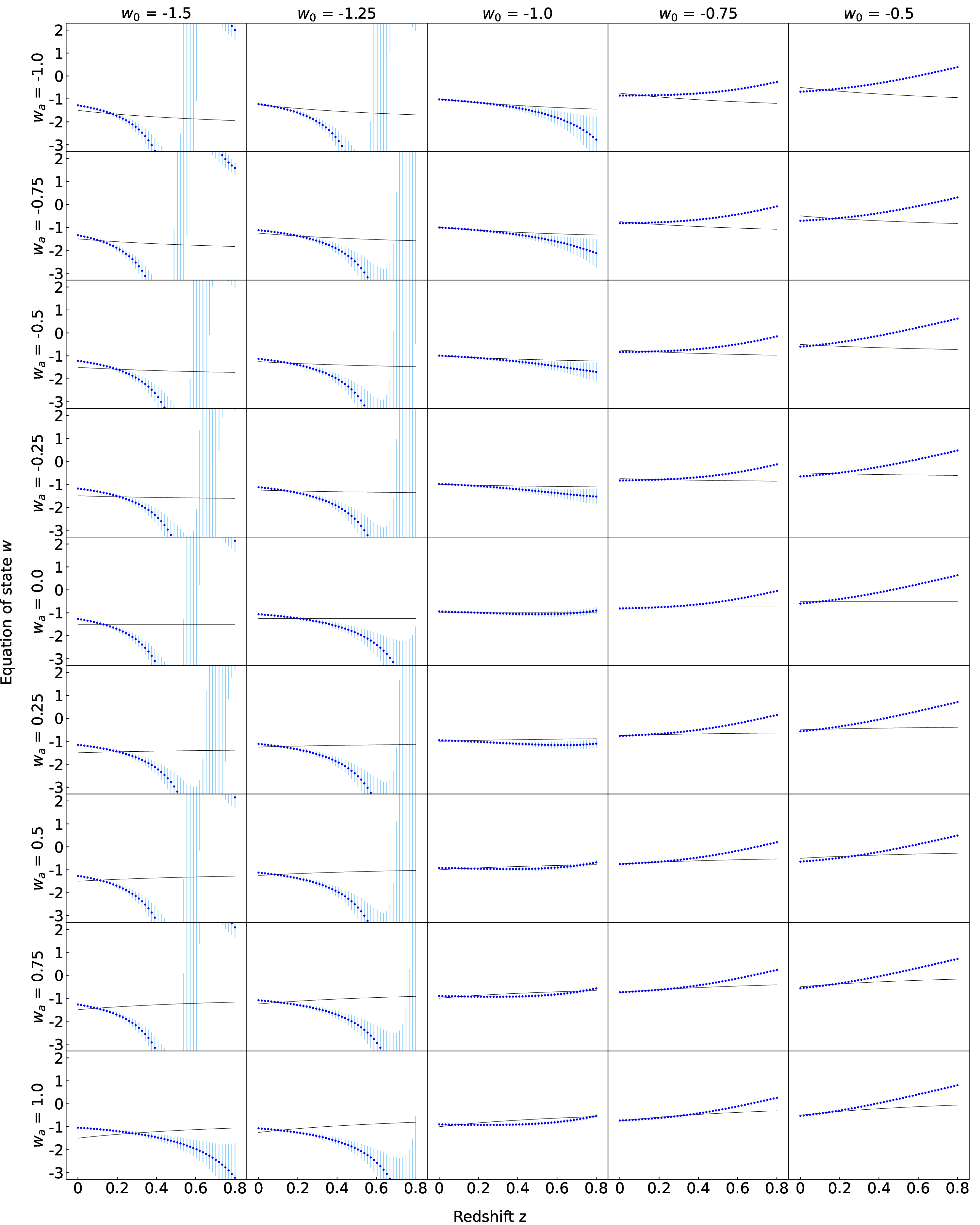}
\caption{\label{fig:w_from_Exponential_generated_SN}Grid of equation of state values $w(z)$ for 10,000 simulated supernovae, where each $w$ is computed using equation \ref{eq:wfinal_methods} applied to an Exponential fit (eq. \ref{eq:Exponential}) to the generated SN data. The plots represent a variety of values of $w_0$ and $w_a$ (CPL equation \ref{eq:CPL}) from -1.5 to -0.5 horizontally, and -1 to 1 vertically, both with a step of 0.25. The black curve represents $w$ from the CPL equation (eq. \ref{eq:CPL}) for the same values. The parameters used to generate the simulated SN were the same as for Fig. \ref{fig:w_from_Quadratic_generated_SN}.}
\end{figure*}

\begin{figure*}
\includegraphics[height=22.5cm]{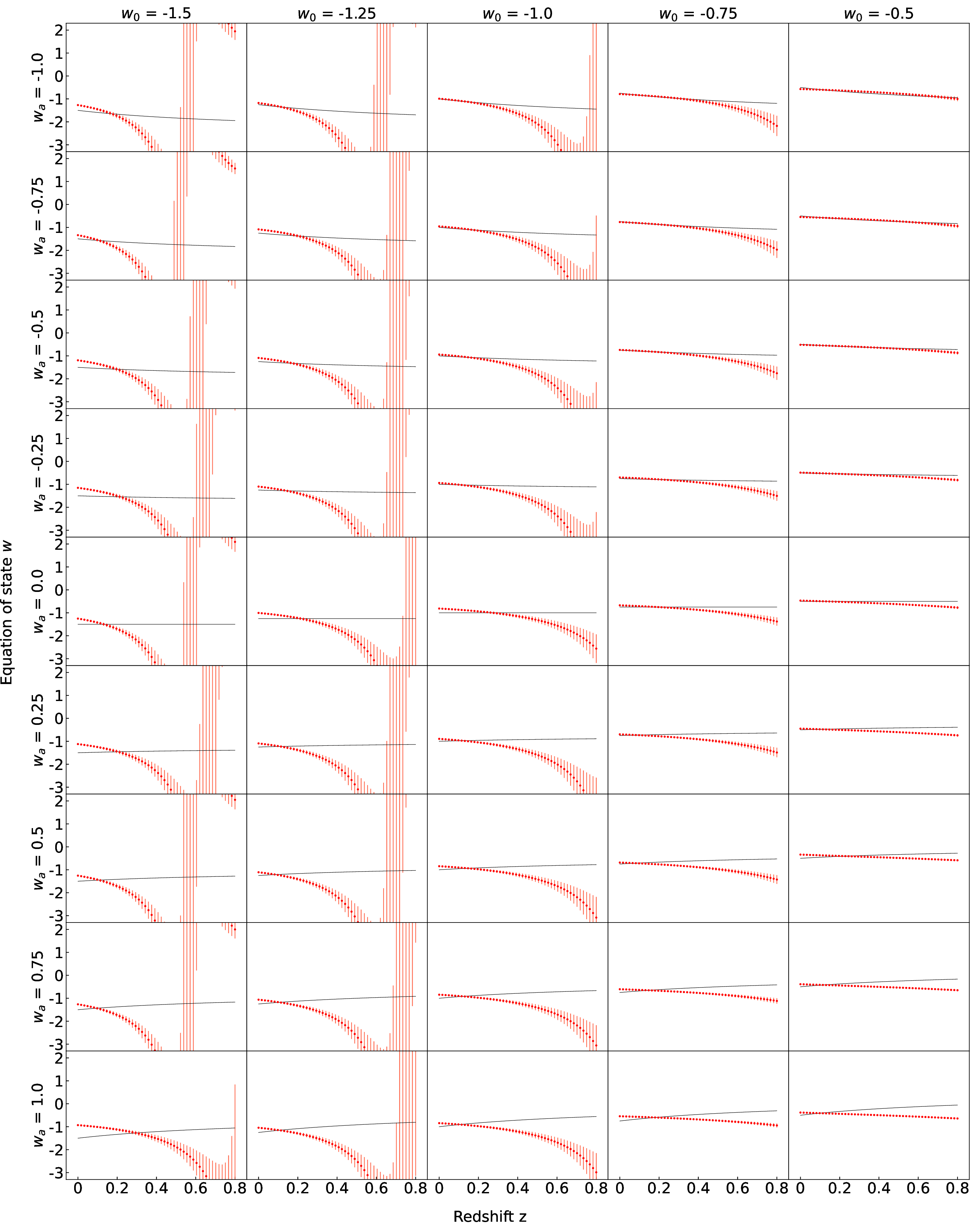}
\caption{\label{fig:w_from_Logarithmic_generated_SN}Grid of equation of state values $w(z)$ for 10,000 simulated supernovae, where each $w$ is computed using equation \ref{eq:wfinal_methods} applied to a Logarithmic fit (eq. \ref{eq:Logarithmic}) to the generated SN data. The plots represent a variety of values of $w_0$ and $w_a$ (CPL equation \ref{eq:CPL}) from -1.5 to -0.5 horizontally, and -1 to 1 vertically, both with a step of 0.25. The black curve represents $w$ from the CPL equation (eq. \ref{eq:CPL}) for the same values. The parameters used to generate the simulated SN were the same as for Fig. \ref{fig:w_from_Quadratic_generated_SN}.}
\end{figure*}

\begin{figure*}
\includegraphics[height=22.5cm]{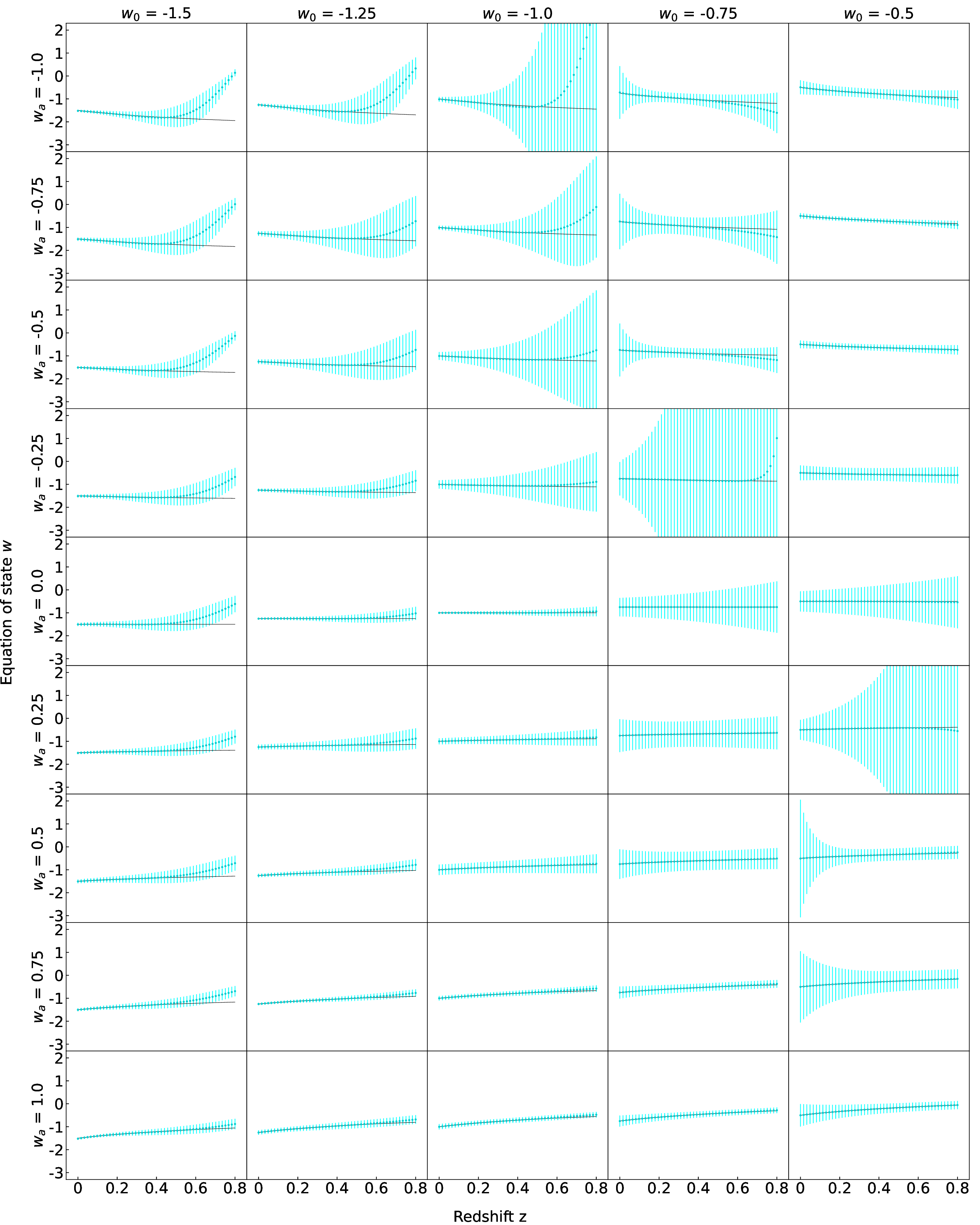}
\caption{\label{fig:w_from_Rational_Fraction_generated_SN}Grid of equation of state values $w(z)$ for 10,000 simulated supernovae, where each $w$ is computed using equation \ref{eq:wfinal_methods} applied to a Rational Fraction (eq. \ref{eq:Rational_fraction}) fitted to the generated SN data. The plots represent a variety of values of $w_0$ and $w_a$ (CPL equation \ref{eq:CPL}) from -1.5 to -0.5 horizontally, and -1 to 1 vertically, both with a step of 0.25. The black curve represents $w$ from the CPL equation (eq. \ref{eq:CPL}) for the same values. The parameters used to generate the simulated SN were the same as for Fig. \ref{fig:w_from_Quadratic_generated_SN}.}
\end{figure*}

\begin{figure*}
\includegraphics[height=22.5cm]{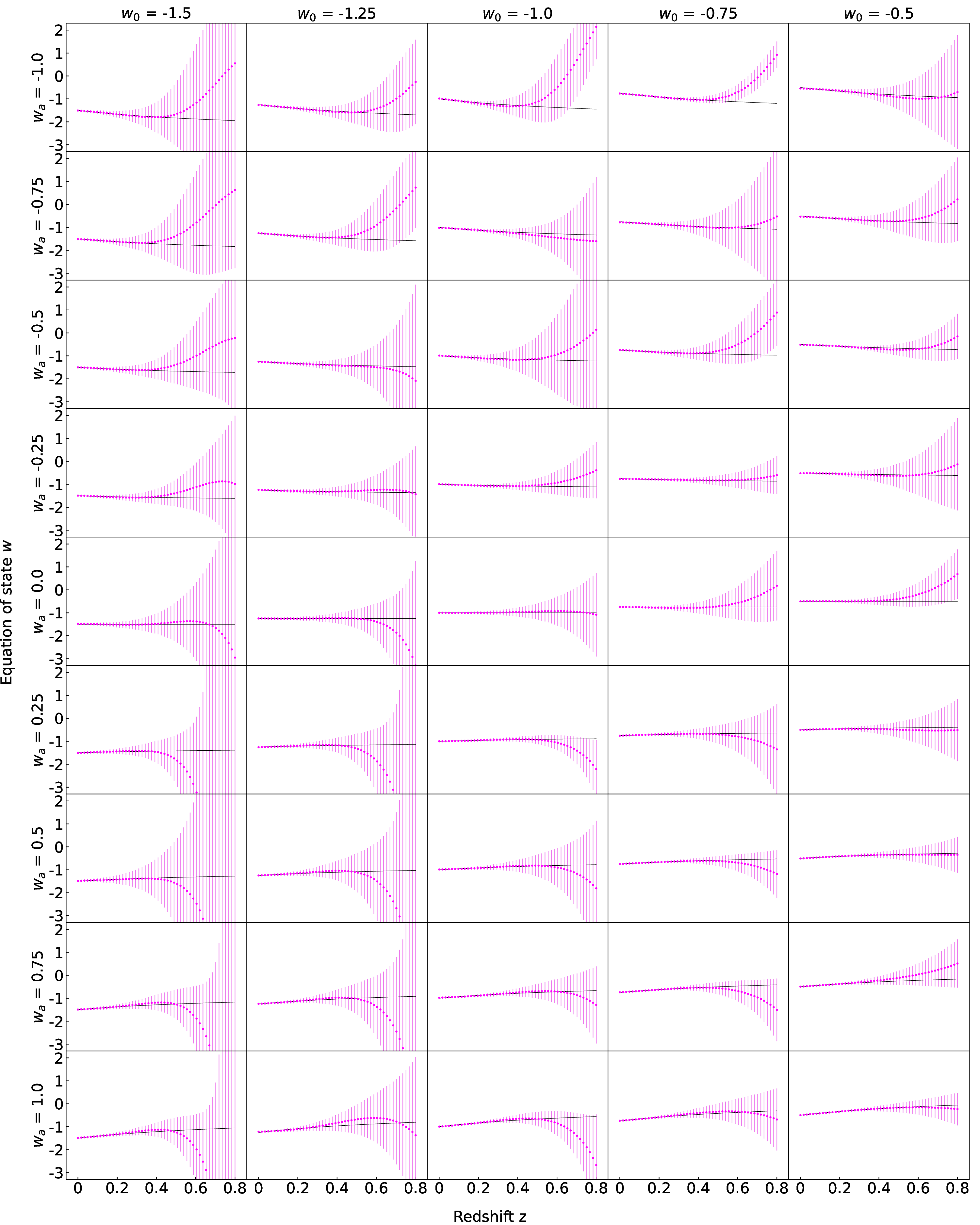}
\caption{\label{fig:w_from_Quartic_1000_generated_SN}Grid of equation of state values $w(z)$ for 1,000 simulated supernovae, where each $w$ is computed using equation \ref{eq:wfinal_methods} applied to a Quartic fit (eq. \ref{eq:Quartic}) to the generated SN data. The plots represent a variety of values of $w_0$ and $w_a$ (CPL equation \ref{eq:CPL}) from -1.5 to -0.5 horizontally, and -1 to 1 vertically, both with a step of 0.25. The black curve represents $w$ from the CPL equation (eq. \ref{eq:CPL}) for the same values. The parameters used to generate the simulated SN were the same as for Fig. \ref{fig:w_from_Quadratic_generated_SN}.}
\end{figure*}

\begin{figure*}
\includegraphics[height=22.5cm]{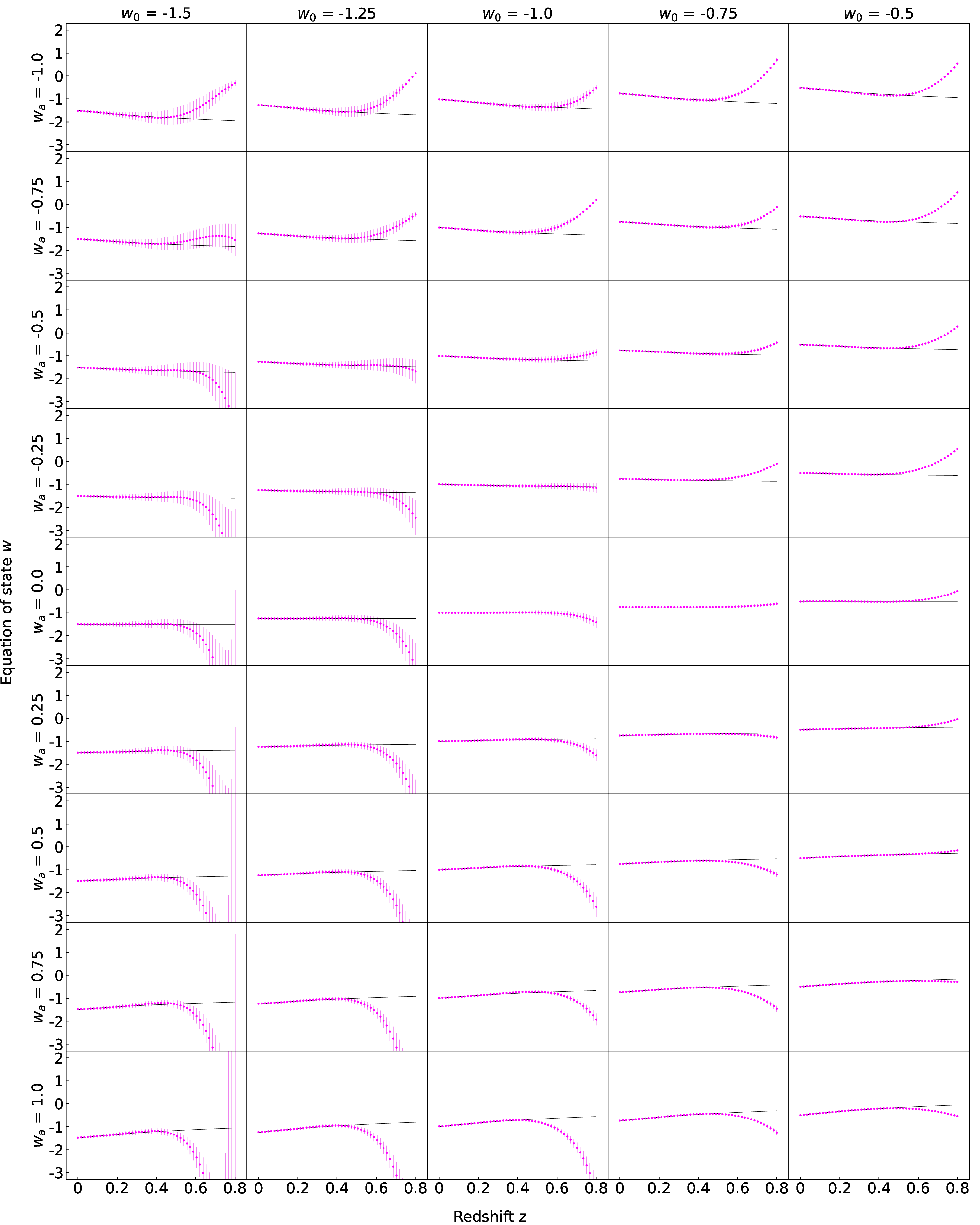}
\caption{\label{fig:w_from_Quartic_1000000_generated_SN}Grid of equation of state values $w(z)$ for 1,000,000 simulated supernovae, where each $w$ is computed using equation \ref{eq:wfinal_methods} applied to a Quartic fit (eq. \ref{eq:Quartic}) to the generated SN data. The plots represent a variety of values of $w_0$ and $w_a$ (CPL equation \ref{eq:CPL}) from -1.5 to -0.5 horizontally, and -1 to 1 vertically, both with a step of 0.25. The black curve represents $w$ from the CPL equation (eq. \ref{eq:CPL}) for the same values. The parameters used to generate the simulated SN were the same as for Fig. \ref{fig:w_from_Quadratic_generated_SN}.}
\end{figure*}

\begin{figure*}
\includegraphics[height=22.5cm]{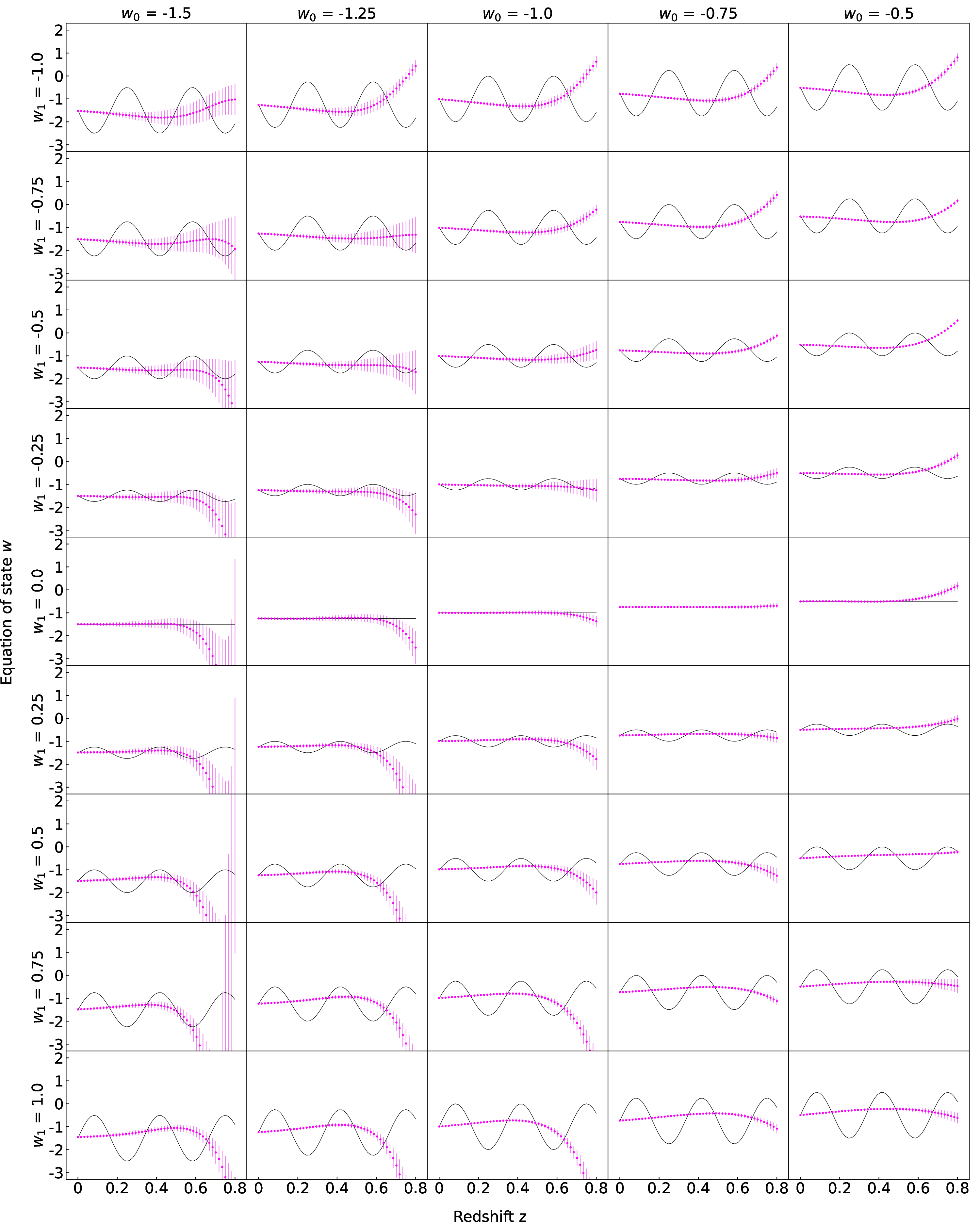}
\caption{\label{fig:sin_w_from_Quartic_generated_SN}Grid of equation of state values for the sine version of $w(z)$ (eq. \ref{eq:sine_variable_w}) for 10,000 simulated supernovae, where each $w$ is computed using equation \ref{eq:wfinal_methods} applied to a Quartic (eq. \ref{eq:Quartic}) fitted to the generated SN data. The plots represent a variety of values of $w_0$ and $w_1$ (CPL equation \ref{eq:CPL}) from -1.5 to -0.5 horizontally, and -1 to 1 vertically, both with a step of 0.25, and $w_2 = 6\pi$. The black curve represents $w$ from equation \ref{eq:sine_variable_w} for the same values. The parameters used to generate the simulated SN were the same as for Fig. \ref{fig:w_from_Quadratic_generated_SN}.}
\end{figure*}

\begin{figure*}
\includegraphics[height=22.5cm]{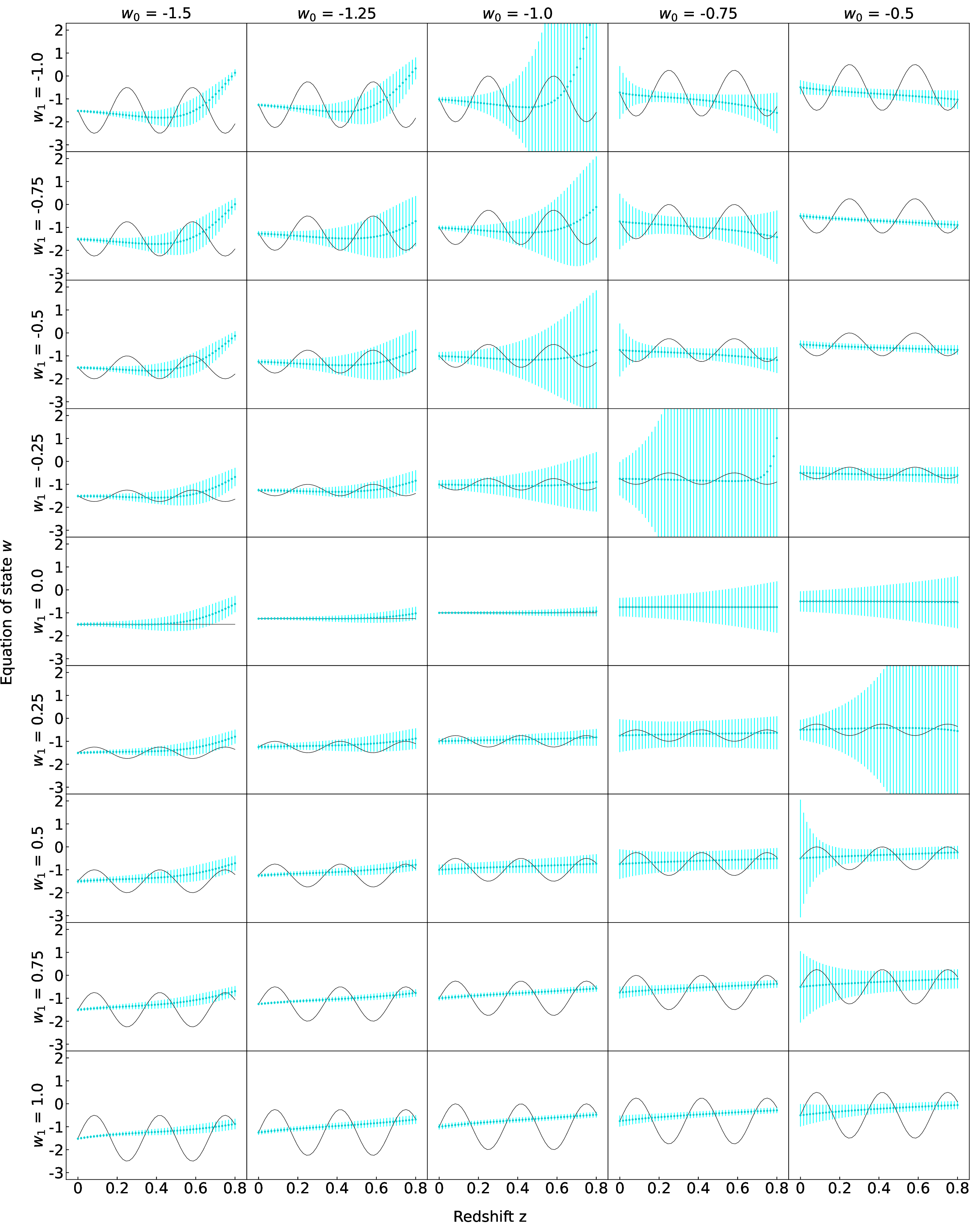}
\caption{\label{fig:sin_w_from_Rational_Fraction_generated_SN}Grid of equation of state values for the sine version of $w(z)$ (eq. \ref{eq:sine_variable_w}) for 10,000 simulated supernovae, where each $w$ is computed using equation \ref{eq:wfinal_methods} applied to a Rational Fraction (eq. \ref{eq:Rational_fraction}) fitted to the generated SN data. The plots represent a variety of values of $w_0$ and $w_1$ (CPL equation \ref{eq:CPL}) from -1.5 to -0.5 horizontally, and -1 to 1 vertically, both with a step of 0.25, and $w_2 = 6\pi$. The black curve represents $w$ from equation (\ref{eq:sine_variable_w}) for the same values. The parameters used to generate the simulated SN were the same as for Fig. \ref{fig:w_from_Quadratic_generated_SN}.}
\end{figure*}

\section{Reconstruction of the equation of state of dark energy from the Pantheon+SH0ES data}

\begin{figure*}
\includegraphics[height=22.5cm]{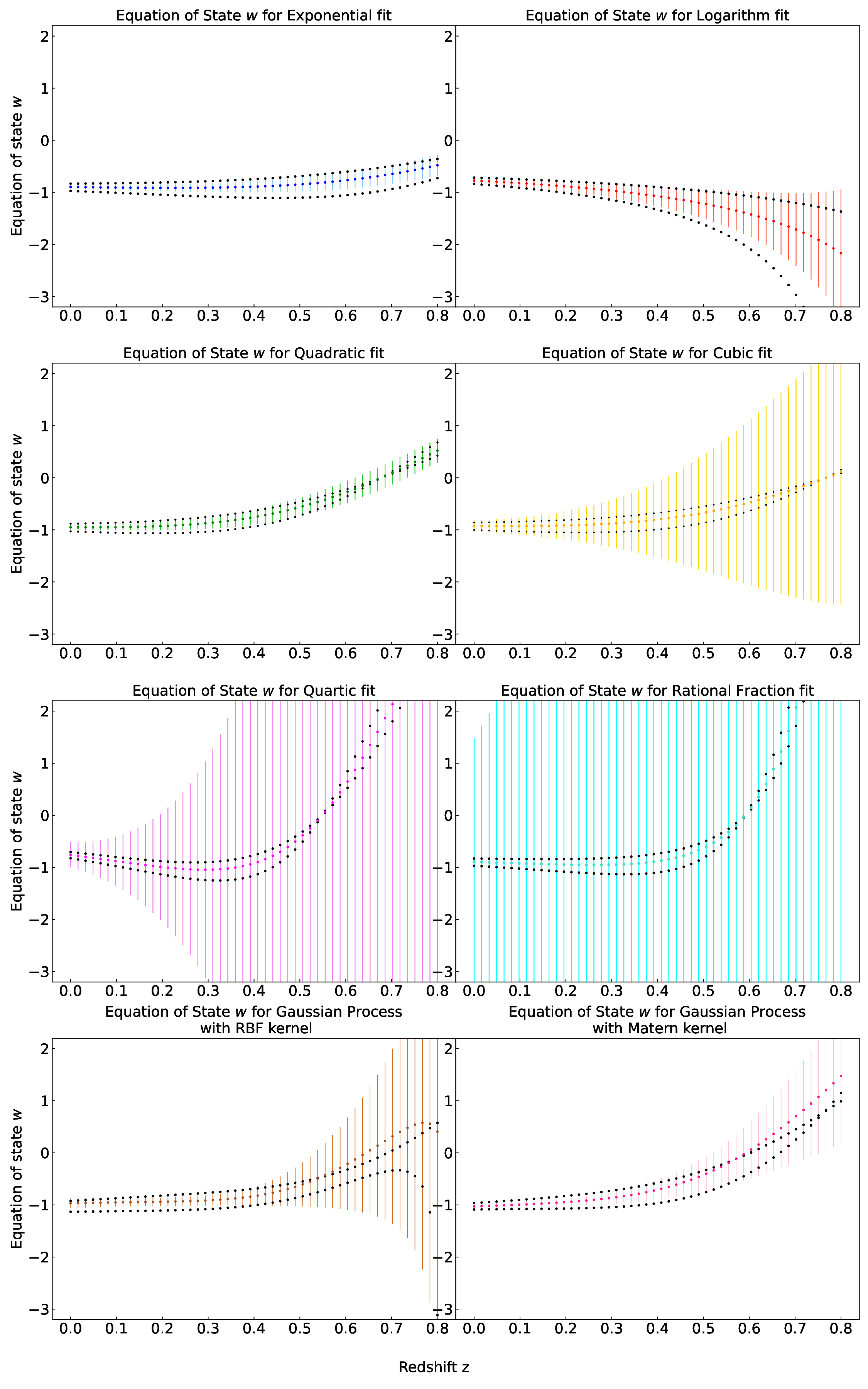}
\caption{\label{fig:All_curves_inc_GP_fitted_to_Pantheon} Equation of state $w$ derived, using equation \ref{eq:wfinal_methods}, from Exponential, Logarithmic, Quadratic, Cubic, Quartic and Rational Fraction curves, plus Gaussian Processes using RBF and Matern kernels, fitted to the Pantheon+ SN data. The plot using the Mat\'ern kernel follows the methodology of \cite{2012JCAP...06..036S}, which is widely used for non-parametric cosmological reconstructions. The black points show the derived values of $w$ for $\Omega_M$ varied by $\pm 0.05$ from the Pantheon+ value of $0.334$ \cite{2022ApJ...938..113S}.}
\end{figure*}

Figure \ref{fig:All_curves_inc_GP_fitted_to_Pantheon} shows the derived equation of state $w$ for all curves (exponential, logarithmic, quadratic, cubic, quartic and rational fraction) fitted to the 1594 SN in the Pantheon+ data set which are not calibrators and where $z < 0.8$, which includes $> 98\%$ of the SN in the data set.
The bottom two panels present the corresponding Gaussian Process (GP) reconstructions for the same subset of 1594 Pantheon+ supernovae, highlighting the impact of kernel choice on the inferred behaviour. In the bottom left panel, a squared--exponential (RBF) kernel with fixed hyperparameters is employed, yielding a highly smooth reconstruction that emphasises the global trend of the data while suppressing small-scale variation. The resulting curve evolves gently with redshift and is accompanied by relatively tight and uniform uncertainty bands, reflecting the strong smoothness prior imposed by this kernel.
The bottom right panel instead adopts a Mat\'ern kernel following the approach of Seikel \& Clarkson \cite{2012JCAP...06..036S}, widely used for non-parametric cosmological reconstructions, 
and which relaxes the assumption of infinite differentiability and permits more local structure in the reconstruction. As a result, the inferred function exhibits greater flexibility, with modestly increased variation and broader, more responsive uncertainty regions that better track the distribution and scatter of the data. Despite these differences, both kernels recover consistent large-scale behaviour, with the Mat\'ern case providing a more conservative representation of the underlying uncertainties.

To allow for uncertainties in $\Omega_M$, we introduced a small variation $\delta_M$. Similarly we introduced $\delta_D$ to vary the dimensionless comoving distance D to incorporate any uncertainty in $H_0$. Therefore the denominator of eq. (\ref{eq:wfinal_methods}) becomes
\begin{equation} \label{eq:denom_with_deltas1}
\Omega_M(1+\delta_M)(1+z)^3{(D'(1+\delta_D))}^2 - 1,
\end{equation}
\begin{equation} \label{eq:denom_with_deltas1}
\approx \Omega_M(1+\delta_M)(1+z)^3{D'}^2(1+2\delta_D)) - 1.
\end{equation}

We introduced a total $\delta= \pm 0.05$ to represent the total of $\delta_M$ and $\delta_D$. This allows for a 15\% variation in $\Omega_M$, much larger than the 5.4\% uncertainty in $\Omega_M$ or the 1.9\% uncertainty in $H_0$, derived from current measurements.

The black points in each panel of figure \ref{fig:All_curves_inc_GP_fitted_to_Pantheon} show the derived values of $w$ with $\Omega_M, D$ varied by a total of $\pm 0.05$. This range is chosen to provide a conservative representation of the combined systematic and statistical uncertainties in the reconstructed parameters, while remaining small enough to probe the local stability of the inferred equation of state.

Non-parametric reconstruction methods, such as Gaussian Processes (GPs), provide a complementary approach to the analytic fits considered above. In this case, they are used to reconstruct the expansion history directly from the data without assuming a specific functional form, offering an alternative perspective on the behaviour of $w(z)$.

Compared to the polynomial and analytic forms considered in this work,
GPs offer a fundamentally different approach to reconstruction. The quadratic, cubic and quartic fits impose a global functional form across the full redshift range, which can lead to biased behaviour: lower-order polynomials lack sufficient flexibility to recover non-trivial evolution, while higher-order forms can introduce artificial structure or instabilities, particularly toward the boundaries of the data. Similarly, exponential, logarithmic and rational functions provide more flexibility in principle, but still constrain the reconstruction to a specific class of behaviour, which can limit their ability to capture unexpected features or lead to increased variance when additional degrees of freedom are introduced.

In contrast, GPs
do not assume a fixed functional form, but instead infer the shape of the function directly from the data through the covariance structure. This allows the reconstruction to adapt locally to the data density, avoiding the global biases inherent in polynomial fits. As seen in figure \ref{fig:All_curves_inc_GP_fitted_to_Pantheon}, the GP reconstructions recover the same broad behaviour as the analytic fits, but with uncertainty bands that more transparently reflect the data quality and sampling. In particular, the GP approach avoids the strong dependence on the chosen fitting function evident in the spread between polynomial and rational reconstructions, providing a more agnostic estimate of $w(z)$.

However, this increased flexibility comes with its own limitations. While the analytic fits provide explicit functional forms that are straightforward to differentiate and interpret, GP reconstructions depend on the choice of kernel and hyperparameters, which effectively act as a prior on the smoothness of the solution.
To quantify the performance of the different fitting functions, we also evaluate the goodness-of-fit using the reduced chi-squared statistic, $\chi^2$, for each model. We find that lower-order polynomials generally provide poorer fits to the simulated data, while higher-order and more flexible forms (such as quartic and rational functions) achieve improved $\chi^2$ values, consistent with their greater ability to capture the underlying structure. However, improvements in goodness-of-fit do not necessarily translate into more accurate reconstructions of $w(z)$, as increased flexibility can also amplify noise in the derivative estimates. This further highlights the trade-off between fit quality and stability in derivative-based reconstructions.

Moreover, as with the polynomial approach, the need to evaluate derivatives amplifies noise, leading to increased uncertainties in $w(z)$ at higher redshift. In practice, the GP results in this work behave similarly to the higher-order and rational fits in terms of uncertainty growth, but offer a more controlled and systematically motivated way of capturing this behaviour. Taken together, this suggests that GPs 
provide a valuable complement to the analytic fits used here, offering a more model-independent benchmark against which the robustness of the reconstructed equation of state can be assessed.

Overall, all reconstruction methods considered here yield results consistent with a cosmological constant, with no statistically significant evidence for deviation from $w = -1$ within the uncertainties of the Pantheon+SH0ES data. Any apparent variation in $w(z)$ remains within the propagated uncertainties and is not robust across different reconstruction approaches.

While a
ll the fits recover $w \approx -1$ up to $z = 0.4$ and diverge to varying degrees for higher $z$
, the uncertainties are very different for each curve fit, with the exponential and quadratic fits having the smallest uncertainties.
\\
While all recover $w$ to some degree, the low number of SN (compared to 10,000 used in the tests) indicates a larger number of SN would be expected to improve the recovery of $w$. The LSST \cite{2024AA...686A..11P, 2025AA...699A..98S} is expected to dramatically increase the number of observed supernovae, by at least two orders of magnitude, 
which will greatly improve the determination of the equation of state. With such a large data set, our approach should enable recovery of the equation of state accurately up to at least $z = 0.8$.

\subsection*{Comparison with Gaussian Process Reconstruction Methods}


Gaussian Process (GP) methods are a widely used non-parametric approach for reconstructing cosmological functions such as $H(z)$ and $w(z)$ directly from observational data. In this framework, the reconstruction is determined by a covariance kernel, which governs the smoothness and correlation structure of the inferred function.

The main advantage of GP methods is their flexibility and minimal reliance on an assumed functional form. However, this flexibility introduces implicit assumptions through the choice of kernel. For example, squared--exponential kernels enforce strong smoothness, potentially suppressing real features, while Matérn kernels allow greater local variation at the cost of increased uncertainty \cite{2013arXiv1311.6678S}. Consequently, GP reconstructions are not fully model-independent and can depend on kernel selection and hyperparameter optimisation.

The hybrid method presented here provides a complementary approach. The MCMC stage constrains key cosmological parameters such as $\Omega_M$ and $H_0$, reducing degeneracies and anchoring the reconstruction. 
This work combines parameter-constrained reconstruction with analytic fitting in a unified framework, providing a controlled alternative to fully non-parametric approaches such as GPs.
Analytic fits to the distance--redshift relation are then used to obtain the derivatives required to evaluate $w(z)$ via eq. (\ref{eq:wfinal_methods}), without assuming a parametric form for the equation of state.

This approach offers several advantages. Constraining $\Omega_M$ and $H_0$ independently mitigates a major source of uncertainty in derivative-based reconstructions. The use of analytic functions provides transparency and allows direct control over the behaviour of the reconstruction. In addition, the method avoids the need to specify a covariance kernel, removing a key source of prior dependence present in GP analyses.

At the same time, the method shares limitations common to all derivative-based approaches. The reconstruction of $w(z)$ remains sensitive to observational uncertainties and to the choice of fitting function, particularly at higher redshifts. Unlike GP reconstructions, which rely entirely on kernel assumptions and adapt their effective complexity directly from the data, analytic fits impose a global functional structure that may not capture more complex or rapidly varying behaviour, as demonstrated in Sec. \ref{Sec:Tests}, while the hybrid method retains a degree of physical anchoring through independent cosmological parameter constraints.

Overall, the hybrid method is complementary to GP-based approaches. While GPs provide a highly flexible non-parametric framework, the method presented here offers a more controlled reconstruction in which cosmological parameters are independently constrained and modelling assumptions are explicit. The consistency between the two approaches, as shown in Fig. \ref{fig:All_curves_inc_GP_fitted_to_Pantheon}, supports the conclusion that current data remain compatible with $w = -1$ within observational uncertainties. This reinforces that current observational constraints are insufficient to distinguish between the reconstruction approaches at the level of functional form, with differences dominated by methodological assumptions rather than data-driven structure.\newline

\section{Conclusions}

In this paper, we introduced a hybrid reconstruction approach for the dark energy equation of state that avoids imposing a direct parametrisation of $w(z)$ itself. Instead, the reconstruction is performed through analytic representations of the distance--redshift relation and its derivatives. The method is therefore semi-parametric: while it reduces dependence on specific dark energy models such as CPL, it remains sensitive to the assumed functional form used to reconstruct $D(z)$.

Tests on simulated supernova data demonstrate that the approach can recover $w(z)$ with reasonable accuracy, particularly when quartic or rational fraction fits are employed. 
The method is called hybrid, because it requires knowledge of $H_0$ and $\Omega_M$, while keeping the equation of state unrestrained.  

Still, while the method does not assume a specific parametric form of $w$ that is then fitted to the data (eg. CPL), the tests shows that the method has problems with detecting a highly variable equation of state. 

The method presented here is based on the distance--redshift relation derived from Type Ia supernovae. In principle, it can be extended to incorporate additional cosmological probes such as baryon acoustic oscillations (BAO) or cosmic microwave background (CMB) constraints. These datasets provide complementary information on the expansion history and can improve constraints on $H(z)$, $\Omega_M$, and the overall shape of $D(z)$. Incorporating such data would be expected to reduce degeneracies and improve the stability of the reconstruction, particularly at higher redshifts. However, combining heterogeneous datasets introduces additional complexities, including the treatment of correlated systematics and differing redshift sensitivities, which are beyond the scope of the present work and are left for future investigation.

Application to the Pantheon+SH0ES dataset indicates consistency with a cosmological constant ($w \approx -1$) up to $z \sim 0.4$, with increasing divergence and uncertainty at higher redshifts due to limited data density.
The results suggest that current observations do not provide compelling evidence for time-dependent dark energy, but they highlight the sensitivity of reconstruction methods to data quality and sample size. Future surveys such as LSST and Euclid, which will deliver orders of magnitude more supernovae, are expected to significantly reduce uncertainties and enable robust tests of dynamical dark energy models. The hybrid method presented here offers a complementary tool for these upcoming datasets and may help to clarify whether the late-time acceleration is truly driven by a cosmological constant or by evolving physics.

\newpage
\appendix
\begin{widetext}
\section{Derivation of the equation of state $w$}\label{derivation_of_w}

Comoving Distance d(z) is given by:
\begin{equation} \label{eq:Comoving}
    d(z) = \frac{c}{H_0}\int_{0}^{z} \frac{dz'}{E(z')},
\end{equation}
where

\begin{equation} \label{eq:Ez}
E(z)=\frac{H(z)}{H_0}   = 
\sqrt{\Omega_M(1+z)^3 + \Omega_R(1+z)^4 + \Omega_K(1+z)^2 + \Omega_\Lambda e^{3\int \frac{(1+w)}{(1+z')}dz'}},
\end{equation}
\end{widetext}
and $\Omega_M, \Omega_R, \Omega_K$ and $\Omega_\Lambda$ are the Matter, Radiation, Curvature and Dark Energy Density parameters respectively, $c$ is the speed of light, $H(z)$ is the Hubble parameter, $H_0$ is the Hubble constant and $z$ is the redshift.

The dimensionless comoving distance D is:
\begin{equation}
D(z) = \frac{H_0}{c} d(z).
\end{equation}

With equation \ref{eq:Comoving}, we can say
\begin{equation} 
D' = \frac{H_0}{c} d' = \frac{H_0}{c}\frac{c}{H_0} \frac{1}{E(z)} = \frac{1}{E(z)},
\end{equation}

and then using \ref{eq:Ez},
\begin{equation} \label{eq:H0_over_Dprime}
D' = \frac{H_0}{H}, 
\end{equation}
where $H = H(z), D' = D'(z)$.

Simplifying equation \ref{eq:Ez}, with $\Omega_K = 0$ and $\Omega_R = 0$, yields
\begin{equation} \label{eq:H_overH0}
\frac{H^2}{{H_0}^2} = \Omega_M(1+z)^3 + \Omega_\Lambda e^{3\int\frac{(1+w)}{(1+z)}dz}.
\end{equation}

Differentiate with respect to $z$ to find
\begin{equation} \label{eq:diff1}
\frac{2HH'}{{H_0}^2} = 3\Omega_M(1+z)^2 + 3\frac{(1+w)}{(1+z)}\Omega_\Lambda e^{3\int\frac{(1+w)}{(1+z)}dz}.
\end{equation}

Combined with equation \ref{eq:H_overH0}, equation \ref{eq:diff1} becomes
\begin{equation}
\frac{2HH'}{{H_0}^2} = 3\Omega_M(1+z)^2 + 3\frac{(1+w)}{(1+z)}\left(\frac{H^2}{{H_0}^2} - \Omega_M(1+z)^3\right),
\end{equation}

Rearranging for $w$, we then have
\begin{equation} \label{eq:w1}
w = \frac{2HH'(1+z) - 3\Omega_M(1+z)^3 H_0^2}{3({H^2} - H_0^2\Omega_M(1+z)^3)} -1.
\end{equation}

Differentiating equation \ref{eq:H0_over_Dprime}
\begin{equation} \label{eq:Hprime}
H' = -\frac{H_0 D''}{{D'}^2},
\end{equation}

and substituting for $H$ and $H'$ in equation \ref{eq:w1} gives
\begin{equation}
w = \frac{-\frac{2}{3}\frac{H_0}{D'}\frac{H_0 D''}{{D'}^2}(1+z) - \Omega_M(1+z)^3 H_0^2}{{\left(\frac{H_0}{D'}\right)^2} - H_0^2\Omega_M(1+z)^3} -1.
\end{equation}

This simplifies to our equation of state
\begin{equation} \label{eq:wfinal}
w = \frac{\frac{2}{3}\frac{D''}{D'}(1+z) + 1}{\Omega_M(1+z)^3{D'}^2 - 1}.
\end{equation}

\end{document}